\documentclass{ametsocV6.1}

\usepackage[most]{tcolorbox}

\newcommand{\itorpe}{TorNet~}
\newcommand{\itorpec}{TorNet,~}

\title{A Benchmark Dataset for Tornado Detection and Prediction using Full-Resolution Polarimetric Weather Radar Data}

\authors{Mark S. Veillette,\aff{a}\correspondingauthor{Mark S. Veillette, mark.veillette@ll.mit.edu} 
James M. Kurdzo,\aff{a} 
Phillip M. Stepanian,\aff{a} 
John Y. N. Cho,\aff{a} 
Siddharth Samsi,\aff{b}\thanks{Siddharth Samsi's contributions to this work were made prior to affiliation with NVIDIA} 
 Joseph McDonald\aff{a}
}

\affiliation{
\aff{a}{Lincoln Laboratory, Massachusetts Institute of Technology, Lexington, Massachusetts}\\
\aff{b}{NVIDIA Corporation, Santa Clara, California}\\
}

\abstract{Weather radar is the primary tool used by forecasters to detect and warn for tornadoes in near-real time.  In order to assist forecasters in warning the public, several algorithms have been developed to automatically detect tornadic signatures in weather radar observations.  Recently, Machine Learning (ML) algorithms, which learn directly from large amounts of labeled data, have been shown to be highly effective for this purpose.  Since tornadoes are extremely rare events within the corpus of all available radar observations, the selection and design of training datasets for ML applications is critical for the performance, robustness, and ultimate acceptance of ML algorithms.  This study introduces a new benchmark dataset, \itorpec to support development of ML algorithms in tornado detection and prediction.  \itorpe contains full-resolution, polarimetric, Level-II WSR-88D data sampled from 10 years of reported storm events.  A number of ML baselines for tornado detection are developed and compared, including a novel deep learning (DL) architecture capable of processing raw radar imagery without the need for manual feature extraction required for existing ML algorithms.  Despite not benefiting from manual feature engineering or other preprocessing, the DL model shows increased detection performance compared to non-DL and operational baselines.  The \itorpe dataset, as well as source code and model weights of the DL baseline trained in this work, are made freely available.}

\begin{document}

\maketitle

%
%
%
\statement
	 Accurate and timely detection of tornadic signatures in radar data enables forecasters to issue timely warnings and preparedness measures, ultimately saving lives and reducing the devastating impact of tornadic storms. Machine Learning (ML) has already been shown to be an effective tool for detecting key signals in radar that can be used to locate and track existing tornadoes.  To further advance state of the art in this area, this study presents a benchmark dataset that can be shared across the research community for development and validation of ML-based algorithms for tornado detection.  This work also provides a baseline deep learning (DL) model that is able to efficiently detect tornadic signatures in radar data.
%
%

%

\section{Introduction}

The study of tornadoes and their detection/prediction is a well-explored area in mesoscale meteorology, with the focus being almost exclusively on meteorological radar.  Multiple iterations of tornado detection algorithms have been implemented in the Weather Surveillance Radar -- 1988 Doppler (WSR-88D) Open Radar Product Generator (ORPG) over the years \citep{ryzhkov++05b,wdtb11,brown+12b}.  Although the current tornadic vortex signature (TVS) algorithm generally performs well from a detection standpoint, it tends to have fairly high false alarm rates, as shown later in this study.  Radar-based tornado detection methods attempt to identify known tornadic signatures using rule-based algorithms,  and in some cases, have been used as training tools for forecasters.
For example, although not currently operational in the ORPG, \citet{ryzhkov++05b} define the tornadic debris signature (TDS) using polarimetric radar data.  The TDS can be used to confirm a tornado is ongoing using WSR-88D data in near-real-time.  However, in many cases, tornadic debris is not lofted into the primary radar resolution volume due to weak tornado strength, lack of debris source, or overshooting of the radar beam due to range via the curvature of the Earth and the atmospheric refractive index \citep{doviak+93}.

Researchers have turned to artificial intelligence (AI) and machine learning (ML) methods in recent years in an attempt to mitigate these issues and better detect tornadoes, especially when traditional algorithms have either failed or had too many false alarms \citep[e.g.,][]{lagerquist++20}.  \citet{cintineo++18} and \citet{cintineo++20} combined numerical weather prediction (NWP) models, satellite, radar, and lightning data to forecast severe weather, first in an empirical way, then followed up in a Bayesian classifier framework.  Their algorithm, ProbSevere v2.0, is now operational in the Multi-Radar, Multi-Sensor (MRMS) system \citep{smith++16}.  More recently, \citet{sandmael2023tornado} utilized random forests to perform tornado detection with polarimetric WSR-88D data at full resolution.  A series of features were designed that extracted statistical values from within an azimuthal shear region, and these features were used to train the random forest model to determine the likelihood of a tornado being present.  In each of these cases, as with many AI/ML algorithms in meteorology, the results are increasingly impressive; however, the raw datasets and models are not always easily accessible to the broader community.

Within the fields of AI/ML, a significant proportion of the effort to develop algorithms and models lies within the process of dataset curation \citep{bishop+06}.  Building datasets is a key task that often leads to the success (or failure) of an AI/ML approach.  Within the AI/ML fields, the concept of benchmark datasets has become increasingly common in order to partially deal with the difficulties that lie within the data curation phase of research \citep[e.g.,][]{olson++17,wu++17,hu++20}.  Benchmark datasets are carefully curated, public data repositories that are shared with the world in order to accelerate the pace of research and development \citep{thiyagalingam++22}.  Instead of spending weeks, months, or even years developing a dataset, researchers are able to start at a baseline to build what they need, whether that involves adding/modifying the dataset, or developing novel techniques to work with the existing dataset.  In fact, one of the key benefits of benchmark datasets is that research teams around the world can benchmark their models and results on the same baseline for fair comparisons.  These datasets are often used as ``challenge problems,'' where research teams are challenged to beat the current best-performing models with a given benchmark dataset \citep{dueben++22,gadepally++22}.

Within the field of meteorology, benchmark datasets are rapidly growing in popularity.  \citet{dueben++22}, who specifically call out the need for benchmark datasets in the atmospheric sciences, acknowledge the incredible size of Earth science datasets and the common lack of direct applicability to existing benchmark image datasets like ImageNet \citep{russakovsky++15} due to much higher dimensionality.  \citet{dueben++22} define a benchmark dataset as either ``scientific'' or ``competition'' types, with the former being more in line with research problems such as what has been described above, and the latter being more about the broader community and non-domain experts.  However, in the authors' experience, some datasets can serve as both \citep[e.g.,][]{veillette++20}, allowing non-domain experts to contribute sometimes innovative new approaches.  \citet{dueben++22} also specify that these datasets should ``grow'' and develop over time as new solutions develop.  Finally, they also define order-1 and order-2 scientific benchmarks, with order-2 defining a ``complete'' benchmark.  \citet{dueben++22} go on to acknowledge several recent benchmark datasets, although this list is constantly growing \citep[e.g.,][]{rasp++20,rasp2023weatherbench,betancourt++21,haupt++21,prabhat++21}.

Several recent publications have summarized relevant approaches to model and algorithm development that could benefit from benchmark datasets.  \citet{chase++22} and \citet{chase++23} offer deep-dive tutorials into a broad range of applications for the topic in the field.  \citet{mcgovern++23} discuss the latest ML techniques in convective weather, summarizing with a discussion on the challenges to develop ML techniques across such a broad range of scales, as well as the need for ``active cross-sector collaboration on testbeds'' for validation.  Convective weather, especially nowcasting \citep[e.g.,][]{gagne++17,han++19,lagerquist++20,cuomo+21}, is one of the most-studied topics within the meteorological ML space.  Within the convective realm, tornadoes and their formation/maintenance/dissipation are an ongoing challenge for forecasters, as evidenced by stagnating lead times and high false alarm rates \citep{noaa21}.

This study introduces a benchmark dataset, called \itorpec for the development of ML algorithms for tornado detection and prediction.  \itorpe includes full-resolution, polarimetric, Level-II WSR-88D data sub-sampled from 10 years of storm reports.  The data are publicly available and will grow with time, meeting most if not all of the \citet{dueben++22} requirements for an order-2 benchmark dataset in the Earth sciences.  The availability of full-resolution imagery provides flexibility to explore and compare a range of different ML (and non-ML) algorithms.  Of particular interest are deep learning (DL) algorithms, e.g, convolutional neural networks (CNNs), which can automatically detect low-level features from high resolution imagery without the need for manual feature engineering.  Construction and training of DL algorithms is typically more complex compared to  non-DL algorithms like random forests, and thus the availability of a benchmark dataset will allow researchers to more rapidly advance state of the art in this space.  As a demonstration, a CNN baseline for tornado detection trained using \itorpe is provided in this work, and this baseline is shown to outperform other common non-DL algorithms.

The paper is organized as follows. Section \ref{s:dataset} provides the rational, sampling procedure, and processing pipeline for creating \itorpe.  ML applications of the dataset are explored in Section \ref{s:ml}, where several ML baselines for tornado detection are developed and compared.  Finally, the application of the CNN baseline model on selected case studies is provided in Section \ref{s:case_studies}.

\section{Dataset Description}\label{s:dataset}

This section provides the rationale and methodologies for creating \itorpec a benchmark dataset for the study of tornadoes containing full-resolution polarimetric radar data.  This dataset encapsulates a broad variety of convective modes and storm types, including active confirmed tornadic storms, pre-tornadic storm evolution, non-tornadic rotating storms, non-rotating severe storms, and non-severe storms. This dataset was constructed to enable two primary research efforts: 1)  Supporting analysis and algorithm development for tornado detection by providing labeled examples of tornadic, rotating non-tornadic, and non-rotating storms; and 2) including time evolution of storms across the continuum of rotation intensities to support tornado prediction by providing potential precursors to tornadogenesis. 

The following subsections provide detail on the structure of the dataset, methodologies for selecting samples, and the data curation process.  Links for downloading \itorpe as well as code samples for working with the data, are provided in the Data Availability section.  

\subsection{Dataset Structure}

\itorpe is comprised of a large number of \textit{samples}, which are the basic unit of the dataset.  Each sample contains a small cropped section, or \textit{chip}, of six WSR-88D variables centered on selected locations $(x_i,y_i)$ and times $t_i$ (described below), where $i$ represents the sample index.  Each variable is provided as an array with axes of range, azimuth, time, and sweep.  All samples in \itorpe are selected from storm \textit{events} found in the National Centers for Environmental Information's (NCEI's) Storm Events Database\footnote{https://www.ncei.noaa.gov/access/metadata/landing-page/bin/iso?id=gov.noaa.ncdc:C00510} (NSED).  Timestamps, or frames, within each sample are assigned a binary label of either ``tornadic'' or ``non-tornadic'' based on whether a confirmed tornado was present at the corresponding timestamp.  Within NSED, storm events are also grouped into storm \textit{episodes}.  In most cases, multiple \itorpe samples are drawn from a given storm event, and, similarly, multiple storm events may be selected from a given storm episode.   While each sample corresponds to a unique chip location and time, it is not uncommon for samples in \itorpe to have some overlap in time and/or space.  

\begin{figure}[h]
\centering
\begin{tcbraster}[raster columns=2,
    raster rows=1,
    colframe=black,
    raster equal height,
    boxsep=0pt,
    left=0pt,
    right=0pt,
    top=0pt,
    bottom=0pt,
    hbox,
    left skip=0pt,
    right skip=0pt,
    boxrule=0.5pt, 
    ]
\tcbincludegraphics{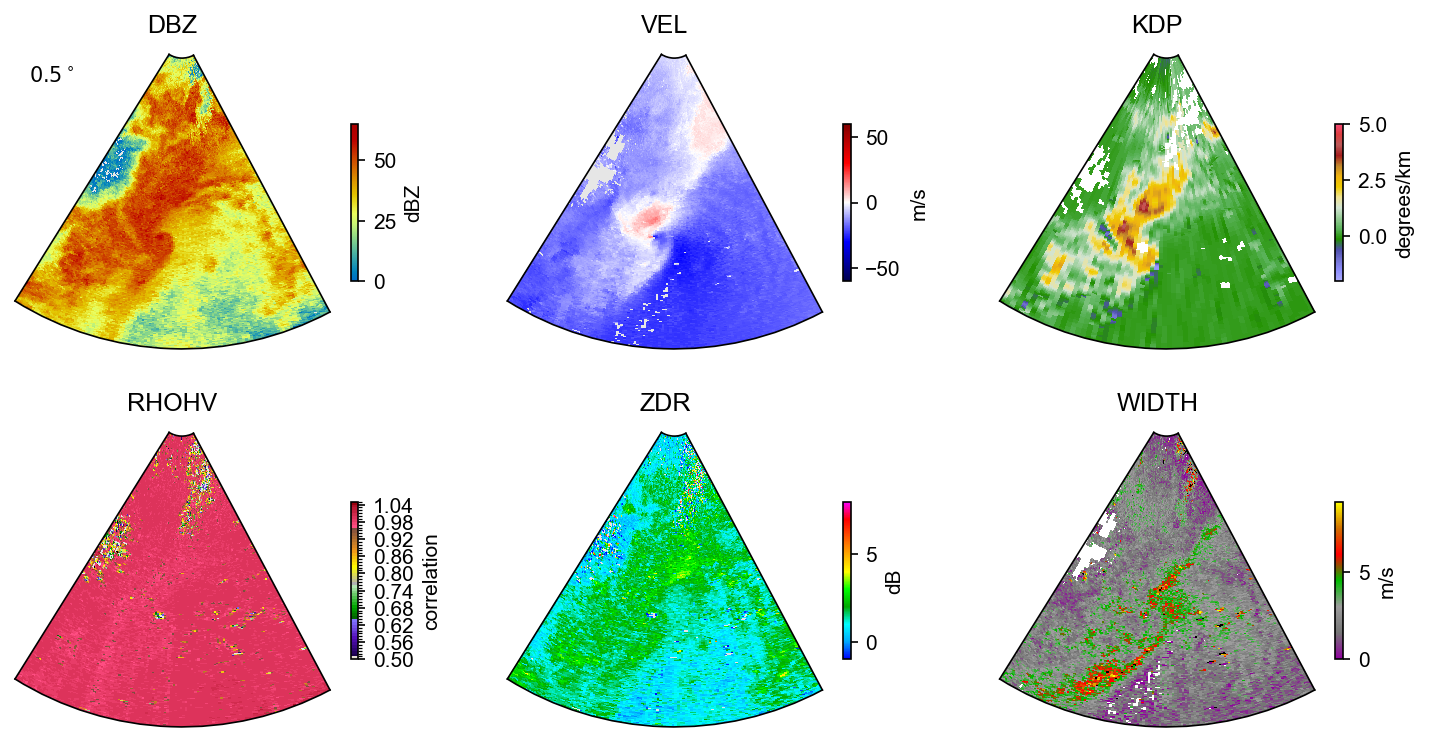}
\tcbincludegraphics{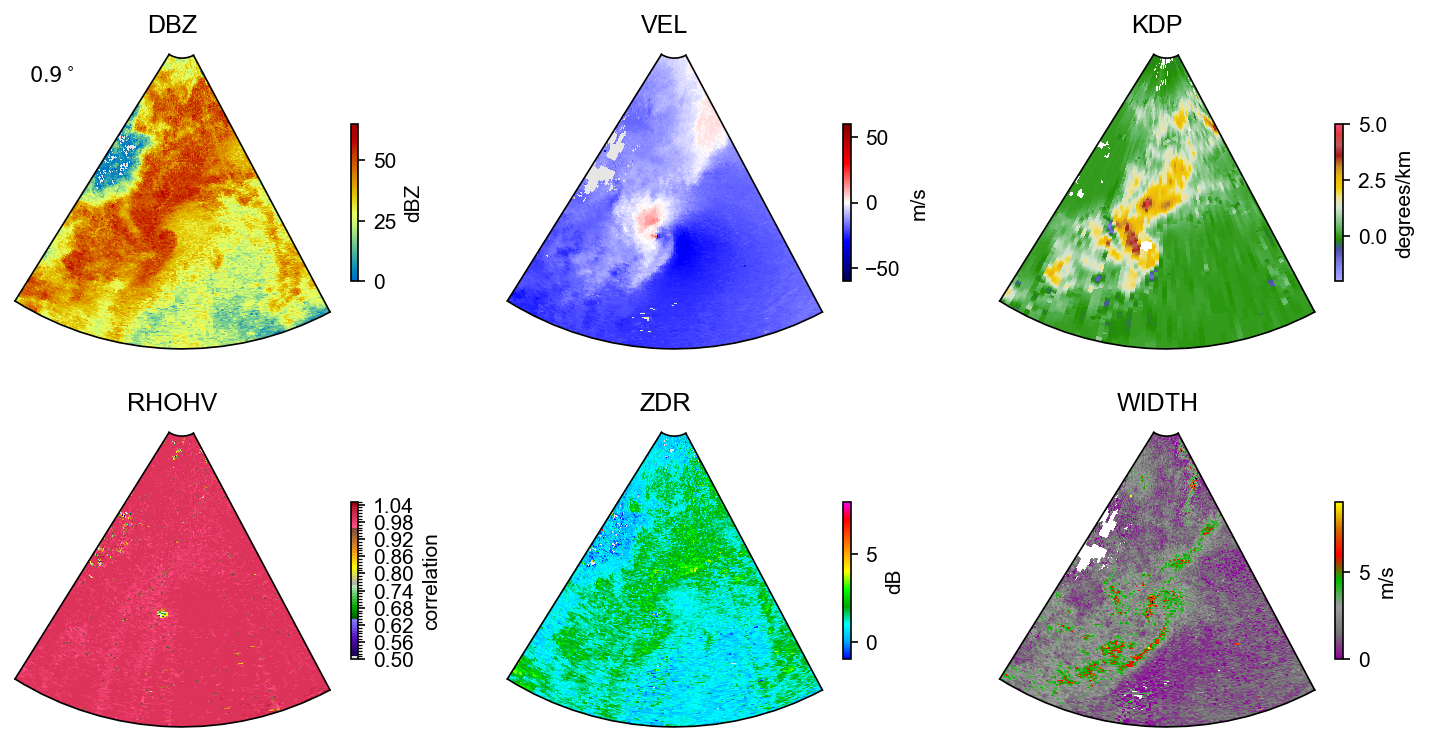}
\end{tcbraster}
\caption{ One sample chip from the \itorpe dataset.  Each sample contains imagery of two elevation sweeps  (0.5$^\circ$ left,  0.9$^\circ$, right) containing six radar variables: Reflectivity factor (DBZ), radial velocity (VEL), specific differential phase (KDP), correlation coefficient (RHOHV), differential reflectivity (ZDR), and spectrum width (WIDTH).  Each sample contains radar variables provided on a 60$^\circ$ by 80-km region (in azimuth and range, respectively) centered over sampled locations and times.  The sample shown depicts one time frame of an EF-3 tornado near KGWX on 
23 February 2019 (NSED event ID 799239). }\label{f:sample_chip}
\end{figure}

Fig. \ref{f:sample_chip} provides an example of one frame from a \itorpe sample.  In this visualization, the data are plotted in Cartesian coordinates; however, it is important to note that the underlying data are represented in polar coordinates (range, azimuth).  The rationale for not resampling data to Cartesian in \itorpe was to preserve the native resolution of the data, especially when observations are close to the radar.


\subsection{Sample Categories}

In general, tornadoes are extremely rare events.  This leads to a class imbalance that is associated with well-known challenges in algorithm design \citep{johnson2019survey,ali2019imbalance,tsagalidis2022exploiting}.   To serve as a useful benchmark for classification and prediction tasks, \itorpe attempts to include as many tornadic samples as possible, while also maintaining a realistic imbalance between tornadic and non-tornadic categories. This ensures that false positive rates can be adequately assessed, and that different strategies for addressing class imbalance (e.g., oversampling the minority class) can be considered.  Moreover, the non-tornadic cases in the dataset should contain a sufficient number of ``difficult'' cases, i.e., non-tornadic cases that exhibit certain features that might lead to false warnings, such as a rotation signature in the velocity field.  Such cases are also quite rare (in general), and not accounting for these may lead to algorithms that over-detect tornadoes.



To achieve a suitable balance of tornadic and non-tornadic cases, samples in \itorpe were selected based on three categories: {\it Confirmed Tornado}, {\it Non-tornadic Tornado Warning}, or {\it Non-tornadic Random Cell}. The definitions of these categories (referred to as ``confirmed,'' ``warnings,'' and ``random'') and their rationale are as follows:

\begin{itemize}

\item {\it  Confirmed Tornado}: These are a selection of storm events from NSED with tornadoes confirmed by storm surveys. These events are gold-standard tornado observations, and are useful for identifying and delineating the radar signatures indicative of existing tornadoes or tornadogenesis.  Samples in this category receive a label of ``tornadic.''

\item {\it  Non-tornadic Tornado Warning}: These are \textit{non-tornadic} events where tornado warnings were issued by a N NWS forecaster.  Samples from this category were added so the dataset contains samples that have high tornado potential (in terms of features in the radar data), but never were confirmed to produce a tornado. These storms are useful for identifying and delineating the radar signatures that are indicative of failed tornadogenesis.  Samples in this category receive a label of ``non-tornadic.''

\item {\it  Non-tornadic Random Cell}: These are a wide selection of all other non-tornadic precipitation cells, ranging from pop-up convective showers to stratiform rain, non-rotating severe thunderstorms, mesoscale convective systems without embedded rotation, etc. These storms are useful for characterizing the broad range of radar signatures associated with precipitation, and delineating these signatures from rotating---and potentially tornadic---storms.  Samples in this category also receive a label of ``non-tornadic.''

\end{itemize}

Some samples in \itorpe contain certain time steps that are both tornadic and non-tornadic, e.g., a case where the tornado forms in the last frame of the sample.  For this reason, each sample has a number of labels equal to the number of time steps in the sample.

\subsection{Event Selection}\label{ss:event_selection}


This section provides detail on how events were selected from each of the categories presented in the previous section.  For each category, the output of this phase is a list that contains sample locations $(x_i,y_i,t_i)$ that contain (1) spatial locations $x_i, y_i$ for each sample centered \textit{near} tornadoes/non-tornadoes; and (2) timestamps $t_i$ of each sample chosen at points within an event.  All sample times in \itorpe were from the period of August 2013 -- August 2022.   

For the confirmed category, the NSED formed the basis of finding tornadic events.  For this category, a crude approximation of location of the tornado was formed by linearly interpolating between the start and end points at one-min temporal resolution.  Points $(x,y,t)$ along this path were recorded for the duration of the tornado.  For the warning category, events were chosen using tornado warning polygons obtained from the Iowa State University warning archive\footnote{https://mesonet.agron.iastate.edu/vtec/search.php} and removing any warnings that coincide with confirmed tornado observations within 30 min.  Warning polygons were reduced into linear ``tracks'' $(x,y,t)$ by defining a linear segment within the polygon that crossed its centroid at the climatological azimuthal angle of tornadic motion.  

For the random category, two methods were used for finding sample locations.  First, severe thunderstorm warning polygons were obtained and filtered to exclude those that were collocated with tornado warnings and/or confirmed tornadoes. The center point of the warning, along with the issue time of the warning, was used as the sample location.  Random locations and times from warning events that were far from warning polygons were also included in this category.


These selection procedures generate a large list of events from the confirmed, warnings, and random categories that form the basis of the \itorpe dataset.  At this stage, the number of warnings and random cases far outnumber confirmed cases.  To reduce this to a final list of sample locations, the list was filtered to prevent cross-contamination among categories and to create a balanced set of categories.  The following considerations were made when sampling storm events cross the three main categories:
 
 \begin{enumerate}
 \item {\it Confirmed}: These are the rarest storms, so we keep all available events.
 \item {\it Warning}: These cases are filtered (+/- 30 minutes) to ensure they do not overlap with any confirmed tornado cases, such that any random storm cell selected throughout the event will not be a confirmed tornado (but may potentially be another co-occurring false tornado). 
 \item {\it Random}: These cases are filtered (+/- 30 minutes) to not overlap with any confirmed or warning events, such that any random storm cell selected throughout the event will be less likely to have tornadic characteristics (i.e., was not associated with any tornado warning, whether confirmed or false). These events are the most common, so we subset the resulting storms to achieve a balanced dataset.
 
\end{enumerate}
 
\begin{figure}[h]
\centering
  \noindent\includegraphics[width=35pc,angle=0]{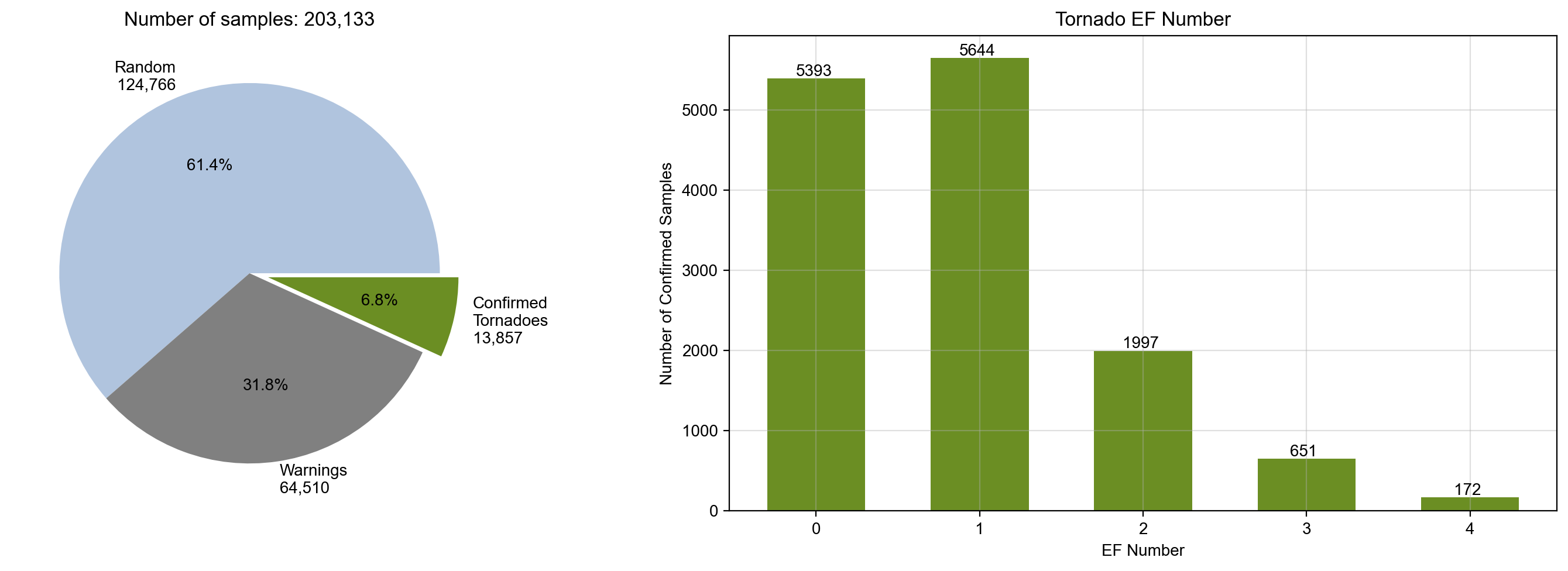}\\
  \caption{ (left) Number of  samples that were created for categories of confirmed tornadoes, non-tornadic tornado warnings, and non-tornadic random cells.  (right) Counts of samples for each EF number.}\label{fig:tornet_summary}
\end{figure}

After performing these steps, the final counts of samples drawn from each category are shown in Fig. \ref{fig:tornet_summary}.  The final dataset contains 203,133 samples, with approximately 6.8\% being from confirmed tornadoes.  Of the non-tornadic cases, approximately one-third are sampled from the warning category.  Fig. \ref{fig:tornet_summary} also breaks down the confirmed tornado category by the  EF number of the associated tornado track, showing that EF-0 and EF-1 tornadoes dominate the confirmed tornado datasets, with decreasing counts for EF-2, 3, and 4.

\subsection{Radar Image Processing}
Given the final list of sample locations, corresponding observations are gathered from the nearest WSR-88D site. The variables extracted for each event include reflectivity factor (DBZ), radial velocity (VEL), spectrum width (WIDTH), differential reflectivity (ZDR), correlation coefficient (RHOHV), specific differential phase (KDP), and the Storm Cell Identification and Tracking information (SCIT; \citet{johnson1998storm}), which is made up of a storm cell's location and unique identifier code. Radar tilts corresponding to $0.5^\circ$ and $0.9^\circ$ elevations for DBZ, VEL, WIDTH, RHOHV, and ZDR were extracted from WSR-88D Level-II files at 0.5$^\circ$ azimuthal resolution were downloaded from Amazon Web Servies (AWS) Open Data Registry.  The SCIT identifiers and KDP fields were obtained from WSR-88D Level-III archives obtained from Google Cloud storage.  

After downloading the required datasets for selected events, various preprocessing steps were performed in order to clean, align, and down-select the data.  Given time $t$, the nearest available data for each variable was located for time $t$, $t-5$ minutes, $t-10$ minutes, and $t-15$ minutes.  For each time, the Level-II radial velocity field is dealiased using the method described in \citep{veillette2023deep}.  KDP, which is provided at 1$^\circ$ azimuthal resolution in Level-III archives, is upsampled to 0.5$^\circ$ using nearest-neighbor interpolation.  All fields are aligned across range and azimuth dimensions. At this stage, all variables are available on axes of [time, sweep, range, azimuth].

The final processing step involves sub-sampling from the range and azimuth dimensions so that dataset samples focus on smaller image chips centered over the locations $(x,y)$ chosen in Section \ref{ss:event_selection}.  This subsampling greatly reduces the size of the dataset, and retains the most meaningful portions of the imagery.  To create a sequence of image chips, the SCIT object closest to the chosen location $(x,y,t)$ is chosen.  Radar variables corresponding to the selected SCIT are obtained, and this data is cropped in the (range,azimuth) dimensions to create a smaller image centered over the selected SCIT.  The size of the chips used in \itorpe were selected to be 80 km (240 250-m gates) by 60$^\circ$ (120 0.5$^\circ$ azimuthal intervals) by two sweeps containing 0.5$^\circ$ and 0.9$^\circ$ tilts.  The spatial dimensions were chosen to be large enough to capture tornado locations that do not closely align with SCIT locations (which is expected to happen in larger storms, especially supercells), or for highly non-linear tornado paths.  The same SCIT location is used for each time step in the sample.




After downsampling, the radar variables in each sample can be represented as a four-dimensional array with shape [T,S,R,A], where $T=4$ corresponds to number of time steps, $A=120$ and $R=240$ are the azimuthal and radial sizes of the chip, and the $S=2$ corresponds to the number of radar sweeps (or tilts) contained in the sample.  Each sample is saved to a separate netCDF file that includes the six radar variables and associated coordinate values.  These files include a variety of metadata, including the sampling category, range-folded masks, event and episode IDs from the the NSED, the EF number of confirmed tornadoes, and information about the NEXRAD site from which data were extracted.





\section{Machine Learning Applications}\label{s:ml}

The \itorpe dataset was designed as a benchmark dataset for ML applications in tornado detection and prediction.  A number of ML tasks can be studied using the dataset, including classification (tornado detection), time-series prediction (forecasting), explainable AI (XAI) methods, automated feature extraction, measures of variable importance, unsupervised learning, and possibly others.  Moreover, \itorpe contains all necessary metadata required for augmenting the radar data with other observations and/or numerical weather prediction (NWP) forecasts.

To demonstrate the utility of \itorpec the remainder of this section will provide a number of baseline classification models for the task of \textit{tornado detection}. In this task, the final frame within \itorpe samples is classified as tornadic or non-tornadic.  We begin by providing the strategy utilized for splitting \itorpe into training and testing partitions, and then describe and compare various classification baselines.


\subsection{Partitioning Data into Training and Testing}

To properly assess hold-out detection performance, \itorpe must be split into training and testing partitions. Any splitting approach needs to avoid \textit{leakage}, which is when testing samples are similar/highly correlated with certain training samples.  For this reason, a naive random splitting approach is {\it not} recommended, as samples often overlap in time and/or space.  A more acceptable methodology for splitting the dataset is by assigning certain intervals of time to training, and other time intervals to testing.  This approach requires choosing optimal lengths of time for performing this partition.  Intervals that are too small risk leakage from occurring at the edges of intervals, whereas intervals that are too long may leave out important seasonal variations from training, and may lead to poorer generalizability and/or less reliable test results.  

The split used in the following attempts to maximize seasonal variability in both training and testing, while minimizing data leakage.  Each sample in the dataset is first associated with the start time $t_e$ of its storm episode ID found in the NSED.   Generally, storm episodes consist of groupings of storm events in a given location and time interval, so that events (and hence individual samples from those events) assigned to different episodes should be sufficiently separated in both time and location.  If $J(t)$ represents the Julian day of the time stamp $t$, defined as number of days since January 1 of the calendar year, samples with $J(t_e) \ (\mathrm{mod} \ 20) < 17$ were considered training data, and samples with $J(t_e) \  (\mathrm{mod} \  20) \geq 17$ are considered test samples.    As a final precaution against cases where storm episodes overlap, any training sample that was within 30 minutes and within 0.25 degrees of latitude/longitude of any test sample were removed from TorNet.  This method  allocated 171,666 (84.5\%) samples into training, and 31,467 (15.5\%) samples into testing.  This method of splitting ensures that both training and testing provide coverage for the entire 10-year span of TorNet.   


\subsection{Baseline Models}\label{ss:baselines}
The baseline models provided include three ML algorithms (logistic regression, random forest, and convolutional neural network), and a non-ML algorithm (tornado vortex signature). Two of the ML models, namely logistic regression and random forest, require a feature extraction step where new features are estimated (e.g., computation of the azimuthal shear field), and full-resolution imagery associated with each sample is reduced to a feature vector of fixed length.  This study implemented a feature extraction approach similar to that described in \citep{sandmael2023tornado}, with some modifications related to input data sources, processing steps, and feature definition.  The details of the feature extraction used in this study are described in Appendix A.  

\subsubsection{Tornado Vortex Signature}

The Tornado Vortex Signature (TVS) is an operational algorithm within the WSR-88D ORPG.  TVS is designed to detect tornadoes within radar data by analyzing various parameters such as radial velocity and reflectivity to identify rotation characteristics associated with tornadoes.  In this study, TVS detections were obtained from the WSR-88D Level-III archives.  For each sample in \itorpec a binary classification of ``tornadic'' is made if a TVS detection lies within 30 km of the center of the chip.  Otherwise, the sample is classified as non-tornadic.



\subsubsection{Logistic Regression}

Logistic regression is a statistical and ML model used for binary classification tasks. It estimates the probability that a given input belongs to a specific category. It models the relationship between the input features and the probability of a sample being tornadic using the logistic function, which ensures that the output is between 0 and 1. The model then uses a threshold to make binary predictions. Logistic regression is a popular baseline because of it simplicity and efficiency.  The class \texttt{LogisticRegression} from Scikit-Learn \citep{pedregosa2011scikit} was used to fit this model.  Hyperparameters \texttt{C} and \texttt{class\_weight} were optimized during the training phase.

\subsubsection{Random Forest}

Random forest is an ensemble learning algorithm used for both classification and regression tasks.  It combines multiple decision trees to make more-accurate predictions. Each tree in the ensemble is trained on a random subset of the data and a random subset of the features. The binary predictions from individual trees are averaged to compute an average prediction score. Random forest is known for its high predictive accuracy, ability to handle large datasets with many features, and resistance to outliers and noise in the data. It is a popular choice in machine learning, especially for cases with tabular datasets.  The class \texttt{RandomForestClassifier} from Scikit-Learn was used to fit this model.  Hyperparameters \texttt{n\_estimators}, \texttt{criterion}, \texttt{max\_depth}, \texttt{min\_samples\_split}, \texttt{min\_samples\_lead} and \texttt{class\_weight} were optimized during the training phase.

\subsubsection{Convolutional Neural Network (CNN)}

CNNs are deep-learning models that have proven to be effective for image classification tasks due to their unmatched ability to handle visual data~\citep{he2016deep}.  The convolutional layers in a CNN apply filters to detect low-level features in the images, while deeper layers progressively learn more complex and abstract features. This hierarchical feature extraction allows CNNs to automatically learn relevant representations from raw pixel data, reducing the need for manual feature engineering.  For this reason, the CNN baseline used in this work only takes raw radar imagery as input, and no additional processing or manual feature extraction is performed to augment the data.  In particular, a derived azimuthal shear field is {\it not} provided to the CNN.

\begin{figure}[h]
\centering
  \noindent\includegraphics[width=38pc,angle=0]{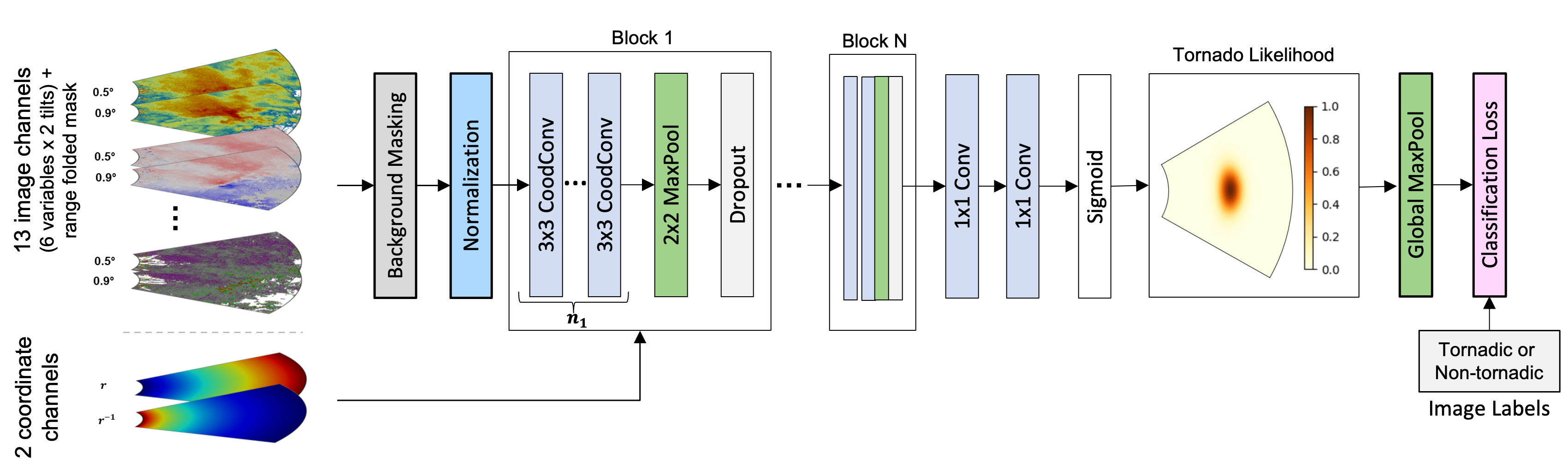}\\
  \caption{Deep learning architecture used for the tornado detection.   Multiple radar modalities shown on the top left are stacked along the channel dimension. Background values are flagged, and channels are normalized to the range [0--1].   The normalized data is processed by multiple VGG-style blocks that utilize CoodConv convolution layers which also ingest the radial coordinates of the radar chip.  The output of the network is an image of tornado likelihoods, which is processed with a global max pooling layer before being compared to the known image labels in the loss function.
   }\label{fig:dl_arch}
\end{figure}

The CNN architecture developed for tornado detection in \itorpe is provided in Fig. \ref{fig:dl_arch}.  The CNN takes as input 13 image channels that include two tilts for each of the six radar variables in \itorpec along with a binary mask of range-folded gates.  In the interest of keeping this baseline simple and portable, only the last frame of each \itorpe sample is provided to the model, although this architecture can easily be extended to take additional frames.  Background gates in each channel are flagged, and the variables are each normalized to the range [0--1], with background and range-folded gates getting a value of -3.  

After masking and normalization, the resulting 13-channel tensor is processed by a series of convolutional blocks, which applies non-linear transformations and halves the spatial resolution of the tensor through pooling.  The number of output channels from Block 1 is a hyperparameter, which is doubled by each subsequent block.  These blocks are similar to convolutional blocks in the well-known VGG19 architecture~\citep{simonyan2014very}; however, instead of traditional convolutional layers, these use a variation called the \texttt{CoordConv} layer introduced in \citet{liu2018intriguing}.  The reason for this modification is because \itorpe data samples are provided on a range-azimuth grid, and thus there is not uniform spatial layout across rows and columns of the input grids (e.g., gates are spatially closer in the azimuthal dimension closer to the radar).  This violates spatial invariance of the convolution assumed in traditional image processing applications.  In order to incorporate knowledge of the coordinate system into the convolution, the \texttt{CoordConv} concatenates additional layers to the input tensor prior to applying convolution, namely a scaled radial coordinate $r$, along with its inverse $r^{-1}$.  The  \texttt{CoordConv} returns the output of the convolution, along with a (possibly downsampled) copy of the input coordinate tensor so that it can be used in downstream \texttt{CoordConv} layers.  The additional inputs to the convolution allow the network to more easily recognize when data are closer to the radar, which is an important factor when recognizing tornadic signatures.

After being processed by the convolution blocks, the set input images are transformed into a tensor with lowered spatial dimension, but a much-higher channel dimension.  This channel dimension is reduced to 1 using two 1x1 convolutional layers, each with linear activation, which is processed with a sigmoid function to map values to [0,1].  The output of the sigmoid is considered a ``tornado likelihood'' because it is not yet calibrated to a probability.  To create the final image classification, the maximum of all likelihoods within the sample chip is computed with a GlobalMax Pooling layer, and it is passed to the loss function in addition to the true image labels (tornadic or non-tornadic). 

The CNN model described above was implemented using the \texttt{tensorflow} deep learning library~\citep{bisong2019tensorflow}.  The loss function used for classification is a binary cross entropy loss function with label smoothing~\citep{zhang2021delving}.  Label smoothing of 10--20\% was found to be effective at stabilizing training by making the loss less sensitive to cases where confirmed samples showed little or no signature is present, and null cases with strong rotation signatures that did not correspond to actual tornadoes.  The CNN described above includes a number of tunable parameters, some of which were set manually due to the time it takes to train the CNN.  These include the number of convolutional blocks ($N=4$), the number of convolutions per block ($n_i=2,3,3,3$), the number of filters output by each block (starting with 32), as well as optimization settings like learning rate, loss weighting terms for tornadic samples, label smoothing, and number of epochs. Some parameters were tuned during the cross-validation procedure described in Section \ref{ss:cv} below.  For the specific values of these, readers should consult implementation provided in the paper's github repository.

\subsection{Baseline Results}

This section provides the training details and test performance of the baseline models on TorNet.  We begin with an overview of the metrics, then describe the process for hyperparameter selection, and finally detail performance on the test set.  Sample visualizations are also provided with classification labels given by the CNN model.

\subsubsection{Metrics of Performance}

With the exception of TVS, which is a binary indicator, each model outputs a continuous value $0 \leq p \leq 1$ indicating a likelihood of there being a tornado in the input imagery.  For scoring, this continuous value is converted to a binary indicator by choosing an appropriate threshold $0 \leq T \leq 1$.  Given a threshold $T$, classifications over \itorpe can be categorized and aggregated into number of true positives (or ``hits'') $H(T)$,  false positives (or ``false alarms'') $F(T)$, false negatives (or ``misses'') $M(T)$, and true negatives (or ``correct rejections'') $C(T)$.   These counts are used to compute the following metrics which are all dependent on the choice of threshold $T$ (which we omit writing for conciseness): Accuracy: $\mathrm{ACC} = (H+C) / (H+C+M+F)$; true positive rate (i.e., probability of detection, recall): $\mathrm{TPR}= H/(H+M)$; false positive rate: $\mathrm{FPR}= F/(F+C)$; success rate (i.e., precision): $\mathrm{SR} = H/(H+F)$; and critical success index: $\mathrm{CSI} = H/(H+M+F)$.   Relationships of probabilistic outputs to frequency of occurrence can be assessed with the Brier Score (BS) given by $\mathrm{BS}=\frac{1}{N} \sum_i (p_i - o_i)$, where $p_i$ is the continuous output of the model and $o_i$ is the binary label.

To visualize several metrics across all potential thresholds, two kinds of performance curves are computed: (i) receiver operating characteristic (ROC) curves, which display TPR on the y-axis and FPR on the x-axis, and (ii) performance diagrams (PD)~\citep{roebber2009visualizing}, which also show TPR on the y-axis, but instead show SR on the x-axis.  Points along these curves represents different choices of threshold $T$ ranging from 0 (everything is a tornado) to 1 (nothing is a tornado).  CSI can also be visualized in a PD since it is a non-linear combination of both TPR and SR.  For both types of curves, the area under the curve (AUC) can be computed to create a single metric of classifier performance which is not dependent on choice of threshold.  The area under the ROC curve will be abbreviated as AUC, while area under (and to the left of) the performance diagram will be abbreviated as AUC-PD.  In order to provide single values for ACC and $\mathrm{CSI}$, the maximum value of these metrics are taken over all thresholds.

\subsubsection{Cross Validation}\label{ss:cv}

The ML models were trained in a two-step process:  
\begin{enumerate}
    \item Using 5-fold cross-validation (CV) to estimate optimal hyperparameters
    \item Retrain the model on the {\it all} training data using optimal hyperparameters
\end{enumerate}
To perform CV, \itorpe was partitioned into five non-overlapping subsets, or folds, with 2013--2014 in Fold 1, 2015--2016 in Fold 2,  2017--2018 in Fold 3, 2019--2020 in Fold 4, and 2021--2022 in Fold 5.   Each model was trained five times using four folds for training, with the remaining fold used for validation.  Optimal parameter settings were found by examining the AUC metric for non-DL models and by using the lowest validation loss for the CNN model.  For logistic regression and random forest models, the \texttt{GridSearchCV} class in Scikit-learn was used to optimally tune hyperparameters for each split described above and AUC was used to select the best parameters for training the model on the entire dataset.  Table \ref{tab:cv_results} shows the performance of all models across the hold-out folds.  The CNN model had the highest average AUC scores across the folds; however, it also had the largest standard deviation which suggests that the training procedure is most sensitive to initial random seeds, outliers in the training data, and possible other factors.  

To address the imbalance of labels in the datasets for training ML models, different class weights were assigned based on the label, category, and EF number of each sample.  For logistic regression and random forest, balanced weighting of the samples, where tornadic samples were weighted higher based on their proportion, showed slight increase in AUC score, but only by less than half a percent.  For the CNN, the sample weights for confirmed, random, and warning categories were all treated as hyperparameters that were tuned on CV partitions.  It was found that weighting tornadoes with EF rating 2 or higher with a weight of 2.0, warning samples with a weight of 0.5, and all others with weight 1.0 improved performance slightly, but again the highest gains were marginal. 

\begin{figure}[h]
\centering
  \noindent\includegraphics[width=25pc,angle=0]{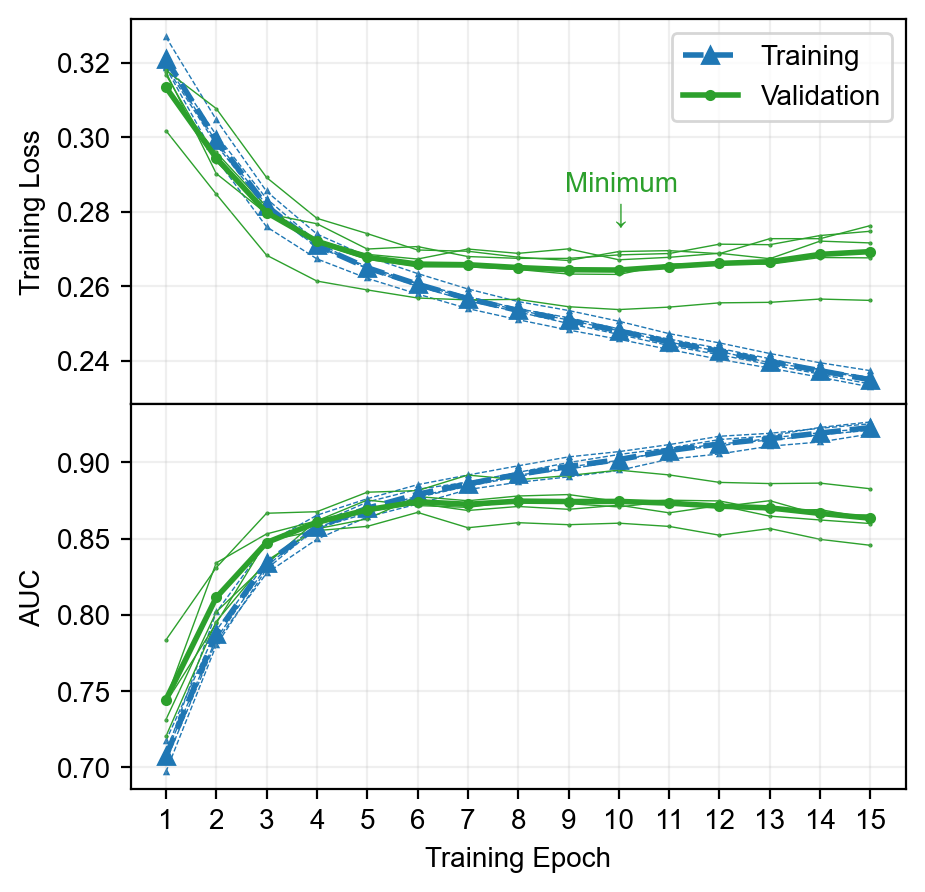}\\
  \caption{ Results of the 5-fold cross validation for the CNN model.  The plots show two metrics of classifier performance as a function of training epoch: training loss (top), and AUC (bottom).  The thin and dashed lines represent the performance of different train/validation partitions, and the solid lines are the average over all folds.  In this case, the optimal number of training epochs (10) was identified by finding the minimum of the training loss on the validation set, indicated in the top plot. }\label{fig:cnn_cv}
\end{figure}

For the CNN model, an important hyperparameter that was identified was the optimal number of epochs to train.  As with all neural networks, training for too long may result in over-fitting, which should be avoided.  Fig. \ref{fig:cnn_cv} shows the cross-validation results across the five training folds for both the training partitions and held-out validation partitions.  Two metrics are shown as a function of training epoch, the training loss (cross-entropy) on top and AUC on the bottom.  In this case, the optimal number of training epochs was selected by minimizing the average validation loss across the five folds (which in this case happened at epoch 10).  It can be seen that for longer training times, the performance of the model on the validation set becomes slightly worse.  Training of the CNN utilized Volta 100 GPUs, and took approximately 2 hours per model that was trained.  
     
\begin{table}
\centering
\begin{tabular}{ | l || rrrrr || rr |}
\hline
 ML Model & Fold 1 & Fold 2 & Fold 3 & Fold 4 & Fold 5 & Mean & Std \\
 \hline \hline
Logistic Regression & 0.8638 & 0.8489 & 0.8432 & 0.8536 & 0.8531 & 0.8525 & 0.0068 \\
Random Forest & 0.8732 & 0.8575 & 0.8511 & 0.8613 & 0.8577 & 0.8602 & 0.0073 \\
CNN & 0.8950 & 0.8744 & 0.8720 & 0.8602 & 0.8708 & {\bf 0.8745} & 0.0114 \\
\hline
\end{tabular}
\caption{Optimal AUC scores of ML models considered from 5-fold cross validation.  The values of each fold represent AUC scores computed on the held-out validation set, and final columns show the average and standard deviation across the five folds.  }\label{tab:cv_results}
\end{table}

\subsubsection{Performance Comparisons}\label{ss:test_performance}

Results are computed for three different views of the testing dataset:

\begin{enumerate}
    \item[(i)]  \textit{All nulls (warnings and random) versus confirmed samples}:   This method utilizes all the testing data.  Because warnings are included, it is expected to have a higher number of false positives as a subset of these samples led to a human forecaster issuing a tornado warning.
    \item[(ii)]  \textit{Randoms versus confirmed samples}:   This partition excludes the more difficult warnings cases, and is expected to reflect performance in situations where a human forecaster did not think any tornadoes were present.
    \item[(iii)] \textit{Warnings versus confirmed samples}: This represents the most difficult view of the data from a classification standpoint, as non-tornadic samples are taken from within tornado warning polygons, and thus are likely to exhibit tornadic features.   
\end{enumerate}  

The performance metrics computed for each model and view are shown in Table \ref{t:test_performance}.  For each  model, this table provides the average performance of the model trained using the entire training set, averaged over five random seeds.  The standard deviation of each metric across the seeds is also shown in parentheses to give a measure of sensitivity of each model to randomness in training.  For ACC, a model that always outputs "No Tornado" obtains an accuracy of 93\% in view (i), which reflects the imbalance of the dataset (and emphasizes that ACC is not a great metric for imbalanced datasets).  The ML models trained on \itorpe all show noticeable gains over (the non-ML) TVS across all metrics considered.  Of the ML models, the CNN is the top performer on average as was seen during the CV.  It is notable that the CNN also carries the highest standard deviation across random initialization, sometimes by an order of magnitude relative to the random forest.  This implies that while the CNN has a higher potential for skill, results are more sensitive to randomness in the training process. Certain strategies, like averaging over ensembles of trained models, or employing more regularization during training, could help alleviate this. 

For view (ii), scores improved overall compared to view (i) due to the removal of the warning category, which confirms that warning samples are indeed more difficult from a detection standpoint.  The ordering of model performance remains virtually the same, with some exceptions (e.g., logistic regression having higher AUC than random forest).   For view (iii), ACC and AUC scores decreased compared to the other cases due to the removal of the random category, while the AUC-PD and CSI scores actually increased slightly relative to (i).  This can be explained by the fact that the majority of samples in the random category result in correct rejections, which do not factor into the calculation of AUC-PD nor CSI, but have a large influence on ACC and AUC.

\begin{figure}[h]
\centering
  \noindent\includegraphics[width=37.5pc,angle=0]{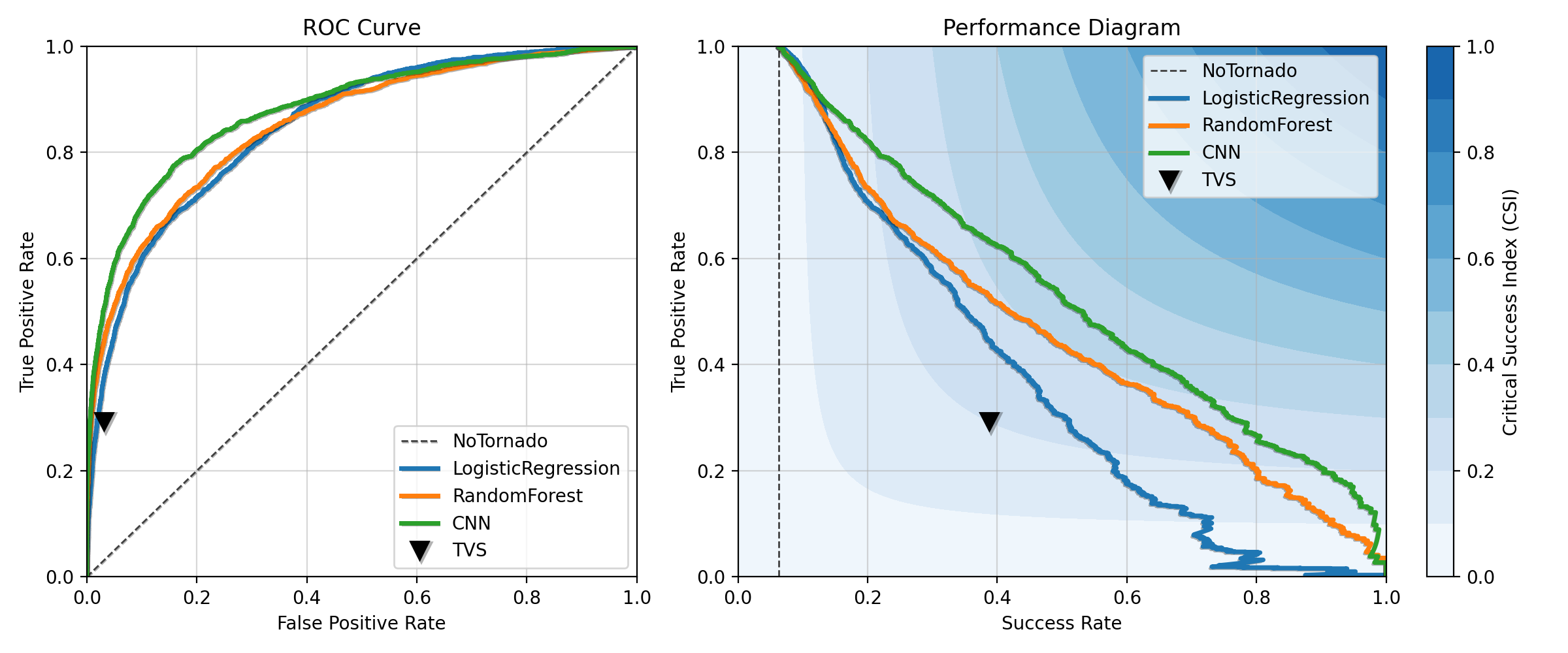}\\
  \caption{ROC curve (left) and performance diagram (right) for the entire test set (confirmed versus all nulls).  In these plots, four baseline models are compared: TVS (which is represented as a single point in this space since it is a deterministic output), logistic regression, random forest, and CNN detectors.  All ML models considerably outperform the TVS algorithm.  In both the ROC and performance diagrams, the CNN model showed the greatest area under the curve (AUC).    }\label{fig:curves_all}
\end{figure}

ROC curves and PD for partition (i) are shown in Fig. \ref{fig:curves_all}.  The ROC curve is shown on the left, and PD on the right, with each displaying performance of the four baseline models, along with the "No Tornado" model (black dashed line).  In both plots, the TVS baseline is indicated by a single point since it is a deterministic model that doesn't depend on any threshold.

In the ROC curve, the TVS baseline falls towards the bottom left portion of the plot, indicating that this algorithm targeted a low FPR (at the expense of also showing low TPR).   In comparison to the ML baselines, the TVS under-performs in regards to both TPR and FPR for a range of different thresholds.  Among the ML models, the CNN offers the best overall AUC, and it shows notable separation from other baselines in the mid-thresholds, but performs similarly to the random forest model for higher thresholds.  A similar trend is noticed in the PD, with the TVS under-performing all ML baselines.   Again, the CNN model shows the highest performance, followed by the random forest and logistic regression models.  ROC curves and PD for partitions (ii) and (iii) of the test set are included in Appendix \ref{s:other_roc}.


\begin{table}[]
\centering
\begin{tabular}{c|cccc|}
\cline{2-5}
\multicolumn{1}{l|}{} & \multicolumn{4}{c|}{(i) Confirmed vs All} \\ \cline{1-5} 
\multicolumn{1}{|l|}{Baseline} &     ACC    & AUC     & AUC-PD     & \multicolumn{1}{c|}{CSI} \\ \hline \hline
\multicolumn{1}{|l|}{NoTornado} & 0.9367 (0) & 0.5000 (0) & 0.0633 (0) & 0.0000 (0) \\
\multicolumn{1}{|l|}{TVS} & 0.9260 (0) & 0.6308 (0) & 0.1583 (0) & 0.2002 (0) \\
\multicolumn{1}{|l|}{Logistic Regression} & 0.9411 (0.0e+00) & 0.8498 (0.0e+00) & 0.3806 (0.0e+00) & 0.2667 (0.0e+00) \\
\multicolumn{1}{|l|}{Random Forest} & 0.9477 (1.3e-04) & 0.8557 (1.1e-03) & 0.4732 (3.3e-03) & 0.3066 (8.7e-04) \\
\multicolumn{1}{|l|}{CNN} & \textbf{0.9499} (2.8e-04) & \textbf{0.8742} (6.2e-03) & \textbf{0.5293} (8.7e-03) & \textbf{0.3380} (6.3e-03) \\
\hline
\end{tabular}

\begin{tabular}{c|cccc|}
\cline{2-5}
\multicolumn{1}{l|}{} & \multicolumn{4}{c|}{(ii) Confirmed vs Random} \\ \cline{1-5} 
\multicolumn{1}{|l|}{Baseline} &     ACC    & AUC     & AUC-PD     & \multicolumn{1}{c|}{CSI} \\ \hline \hline
\multicolumn{1}{|l|}{NoTornado} & 0.9041 (0) & 0.5000 (0) & 0.0959 (0) & 0.0000 (0) \\
\multicolumn{1}{|l|}{TVS} & 0.9272 (0) & 0.6436 (0) & 0.3163 (0) & 0.2783 (0) \\
\multicolumn{1}{|l|}{Logistic Regression} & 0.9356 (0.0e+00) & 0.9161 (0.0e+00) & 0.6562 (0.0e+00) & 0.4538 (0.0e+00) \\
\multicolumn{1}{|l|}{Random Forest} & 0.9404 (1.7e-04) & 0.9090 (6.5e-04) & 0.6969 (1.2e-03) & 0.4598 (1.7e-03) \\
\multicolumn{1}{|l|}{CNN} & \textbf{0.9470} (1.7e-03) & \textbf{0.9206} (5.9e-03) & \textbf{0.7471} (1.2e-02) & \textbf{0.5244} (1.7e-02)     \\ \cline{1-5} 
\end{tabular}

\begin{tabular}{c|cccc|}
\cline{2-5}
\multicolumn{1}{l|}{} & \multicolumn{4}{c|}{(iii) Confirmed vs Warnings} \\ \cline{1-5} 
\multicolumn{1}{|l|}{Baseline} &     ACC    & AUC     & AUC-PD     & \multicolumn{1}{c|}{CSI} \\ \hline \hline
\multicolumn{1}{|l|}{NoTornado} & 0.8432 (0) & 0.5000 (0) & 0.1568 (0) & 0.0000 (0) \\
\multicolumn{1}{|l|}{TVS} & 0.8247 (0) & 0.6082 (0) & 0.2329 (0) & 0.2076 (0) \\
\multicolumn{1}{|l|}{Logistic Regression} & 0.8569 (0.0e+00) & 0.7335 (0.0e+00) & 0.4341 (0.0e+00) & 0.2880 (0.0e+00) \\
\multicolumn{1}{|l|}{Random Forest} & 0.8714 (3.8e-04) & 0.7611 (1.2e-03) & 0.5148 (2.0e-03) & 0.3242 (2.8e-03) \\
\multicolumn{1}{|l|}{CNN} & \textbf{0.8780} (6.3e-04) & \textbf{0.7928} (7.3e-03) & \textbf{0.5689} (7.9e-03) & \textbf{0.3548} (8.7e-03)     \\ \cline{1-5} 
\end{tabular}

\caption{Performance results across baselines models on different views of the \itorpe test set.  The values represent average over each model trained with five different random seeds, with standard deviations in parentheses.}\label{t:test_performance}
\end{table}

\subsubsection{Probabilistic Calibration}
A common post-processing step for classification is calibration, where the output of the ML model is modified so that it better reflects probabilities of observing outcomes \citep{nixon2019measuring}.   It is important to note that the term ``probability'' in this setting reflects the distribution of samples selected for the \itorpe dataset, which is heavily biased towards observing a tornado compared to what a forecaster would observe in the field.  Nevertheless, it is still useful to examine how tornado likelihoods are related to observed frequencies within.  This section only provides calibration results for the CNN model, but a similar analysis can be performed for the other ML baselines.  

\begin{figure}[h]
\centering
  \noindent\includegraphics[width=20pc,angle=0]{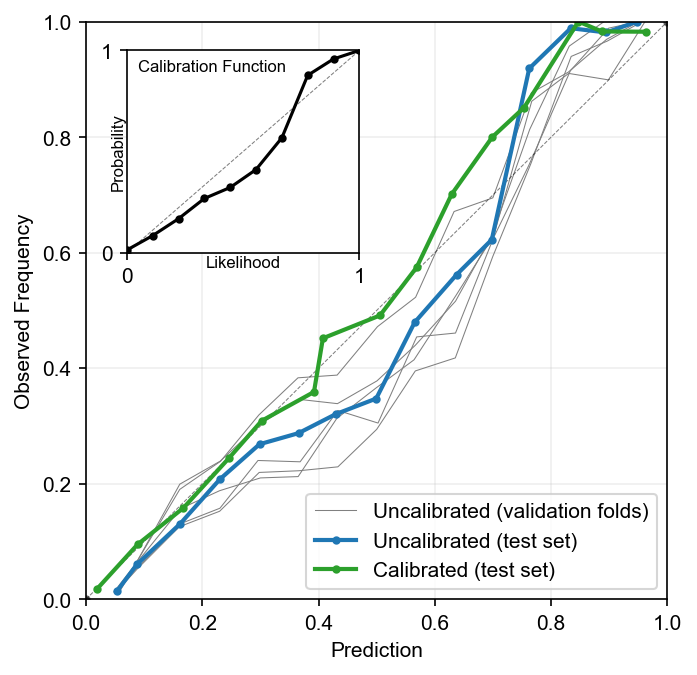}\\
  \caption{ Classifier calibration results.  The grey curves show the comparison between tornado likelihoods computed for a 5-fold CV against observed probabilities, and the blue curve shows comparison on the testing set.  Using the results of the CV, an isotonic regression was fit to calibrate the likelihoods to a probability of observed occurrence, which is the black curve shown in the inset on the top left.  The green curve shows the result of calibration on the test set which better aligns with the 1-1 line. }\label{fig:calibration}
\end{figure}

As a first step, a comparison of likelihood to observed frequency was performed for the five folds of the cross validation study provided in Section \ref{ss:cv}.  The gray curves shown in Fig. \ref{fig:calibration} reflect calibration curves for each of the five folds considered.  Each shows reasonable similarity to the black 1-1 line representing exact calibration, with uncalibrated likelihoods tending to be over-confident in the mid-range (0.2 - 0.7), and under-confident for high likelihoods (>0.8).  The blue line shows a similar result for the test set, which is consistent with the results of the CV.

Using the five calibration curves computed over the CV folds, a calibration function was fit to estimate the observed probability from the predicted likelihood produced by the CNN models.   To fit a calibration function, isotonic regression was used, which estimates a monotone function for inputs in [0,1].  The shape of this fit curve is shown in the inset of Fig. \ref{fig:calibration} in the upper left.  This curve shows the expected pattern of decreasing over-confident mid-level likelihoods, and increasing higher likelihoods.  The green curve in the plot shows the result of calibrating the likelihoods computed over the test set, and shows better alignment with the 1-1 line.  Calibrating the likelihood shows a slight improvement (decrease) in Brier Score, with $\mathrm{BS}=0.0423$ before calibration and $\mathrm{BS}=0.0404$ after calibration.

\subsubsection{Visualizations}

This section provides visualizations of various samples in \itorpe accompanied by the corresponding calibrated probabilistic output $p_D$ of the CNN classifier.  All samples visualized in this section are selected from the test set and were unseen during model training.  Four visualization candidates were randomly chosen from low-, mid-, and high-probability ranges for each of the four categories: (i) hits, where the sample is a confirmed tornado and $p_D>0.5$; (ii) correct rejections, where there is no confirmed tornado and $p_D<0.5$; (iii) misses, where this is a confirmed tornado and $p_D<=0.5$; and (iv) false alarms, where there is no confirmed tornado and $p_D>=0.5$.  Only reflectivity factor and radial velocity are shown; however, the CNN model utilized all variables in the \itorpe samples.

\begin{figure}[h]
\centering
\begin{tcbraster}[raster columns=2,
    raster rows=2,
    colframe=black,
    raster equal height,
    boxsep=0pt,
    left=0pt,
    right=0pt,
    top=0pt,
    bottom=0pt,
    hbox,
    left skip=0pt,
    right skip=0pt,
    boxrule=0.5pt, 
    ]
\tcbincludegraphics{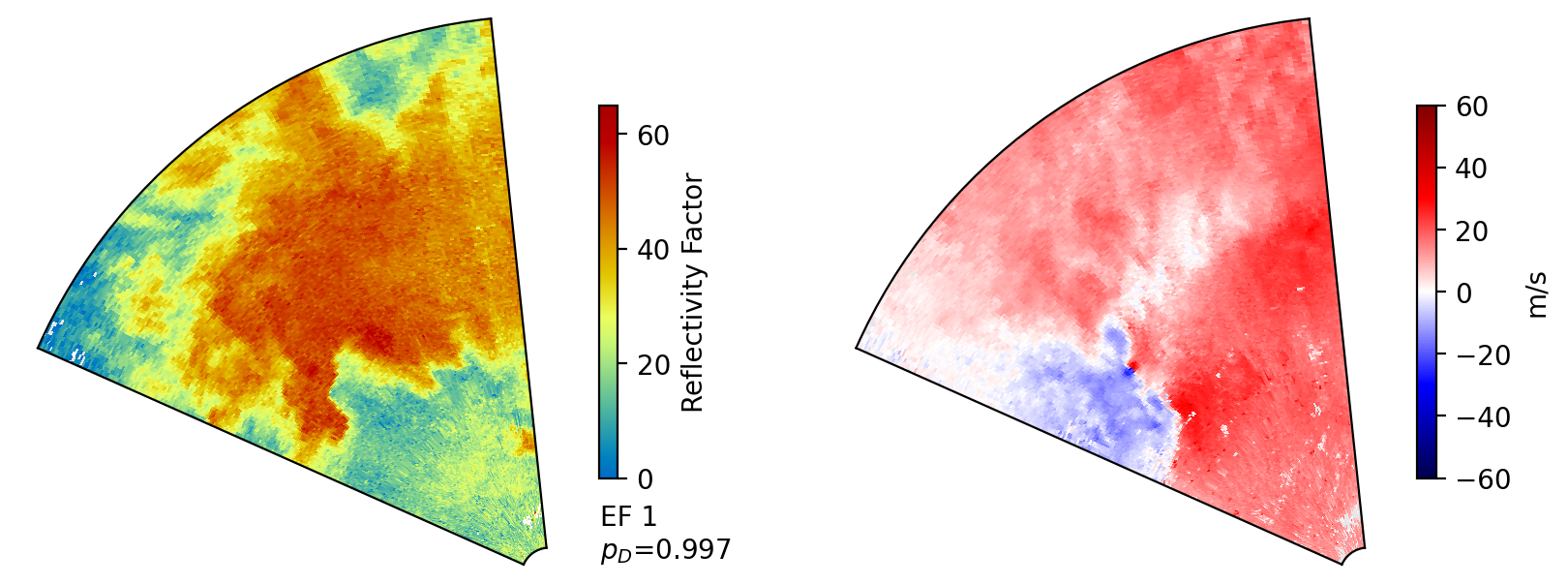}
\tcbincludegraphics{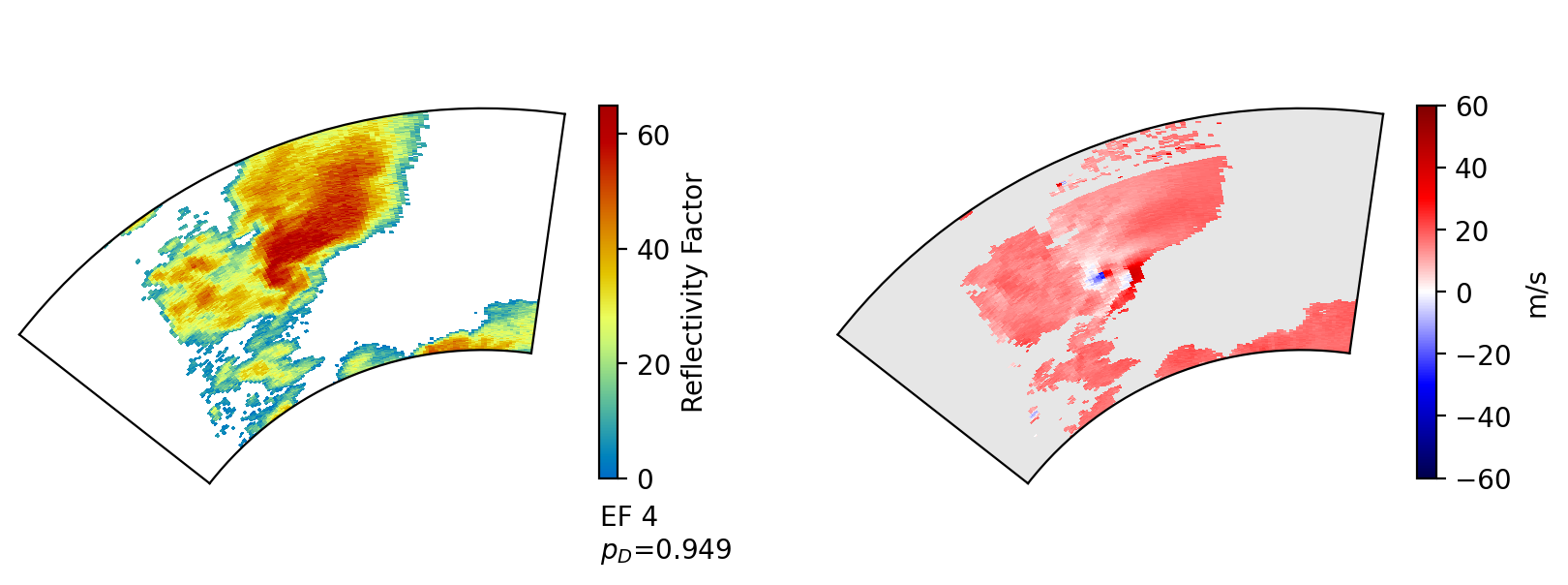}
\tcbincludegraphics{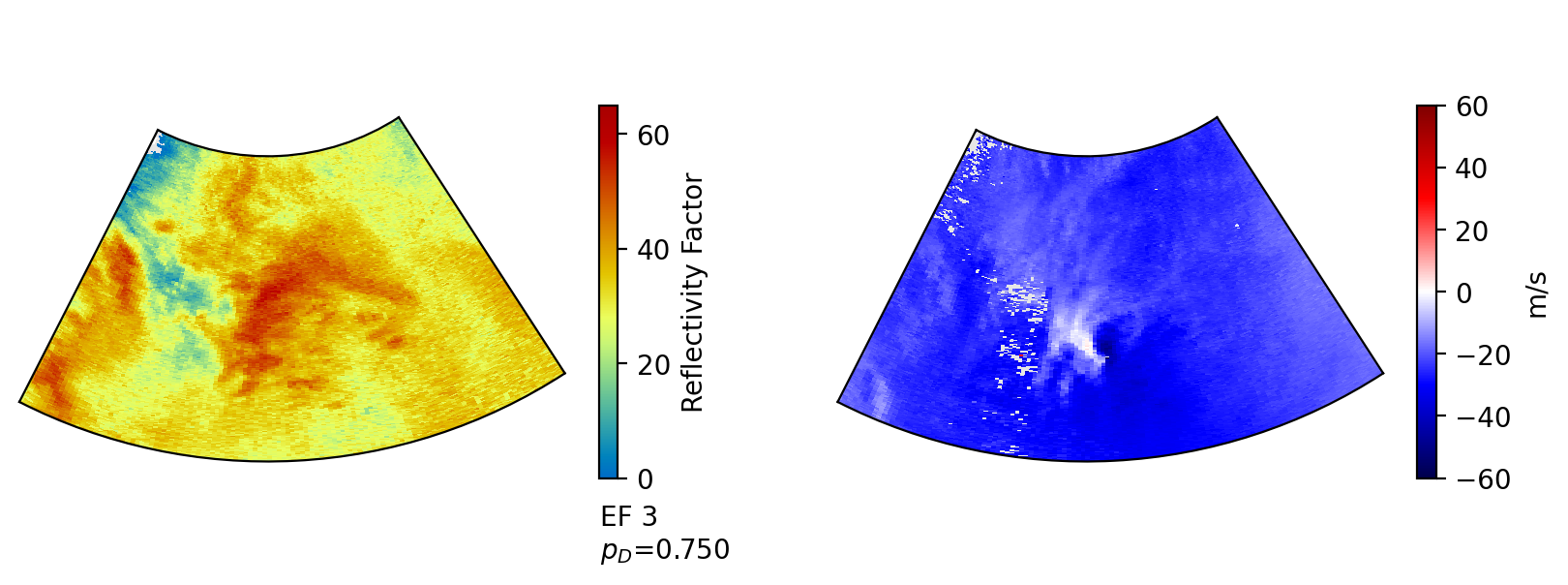}
\tcbincludegraphics{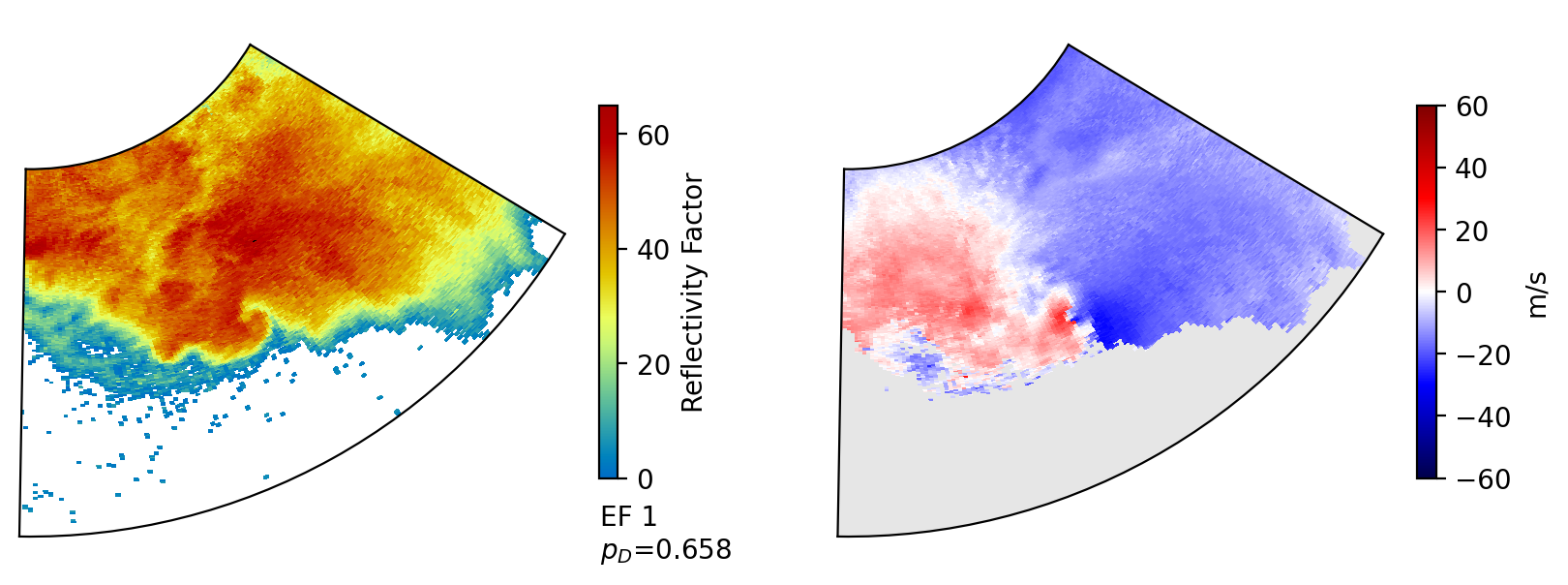}
\end{tcbraster}
\caption{\textbf{Hits}: Example tornadic cases that the CNN model correctly classified.}\label{f:hits}
\end{figure}

\begin{figure}[h]
\centering
\begin{tcbraster}[raster columns=2,
    raster rows=2,
    colframe=black,
    raster equal height,
    boxsep=0pt,
    left=0pt,
    right=0pt,
    top=0pt,
    bottom=0pt,
    hbox,
    left skip=0pt,
    right skip=0pt,
    boxrule=0.5pt, 
    ]
\tcbincludegraphics{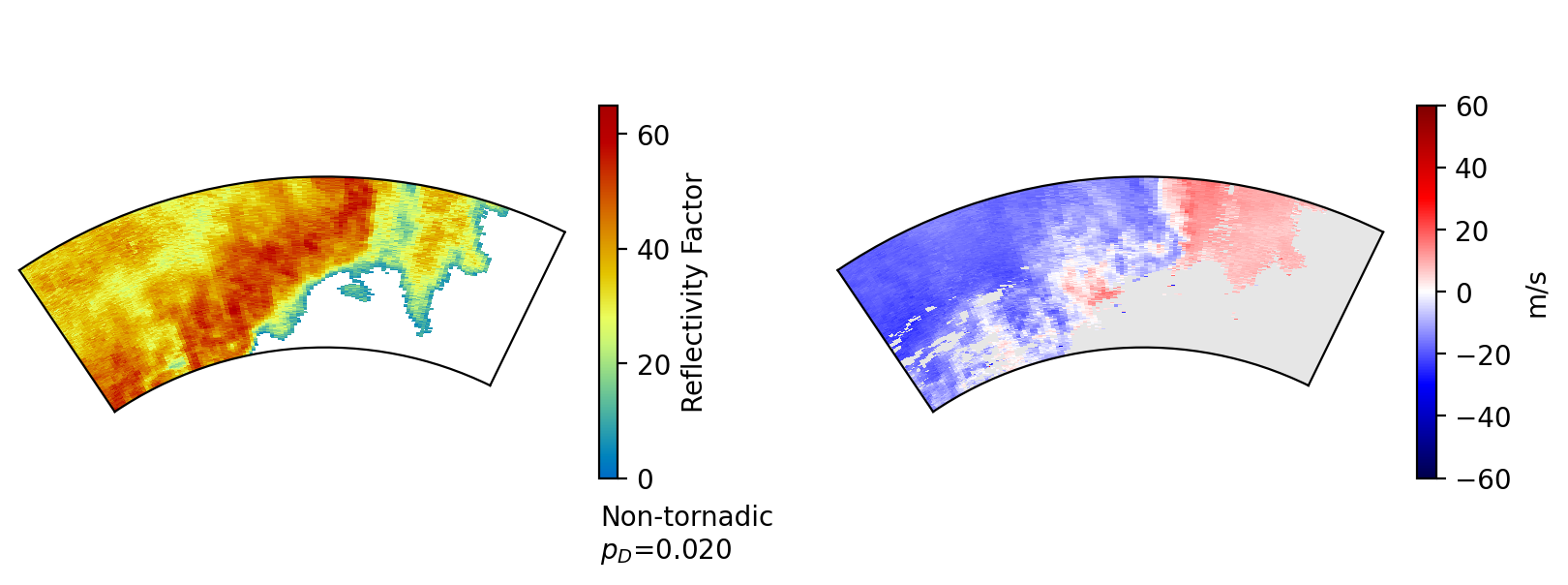}
\tcbincludegraphics{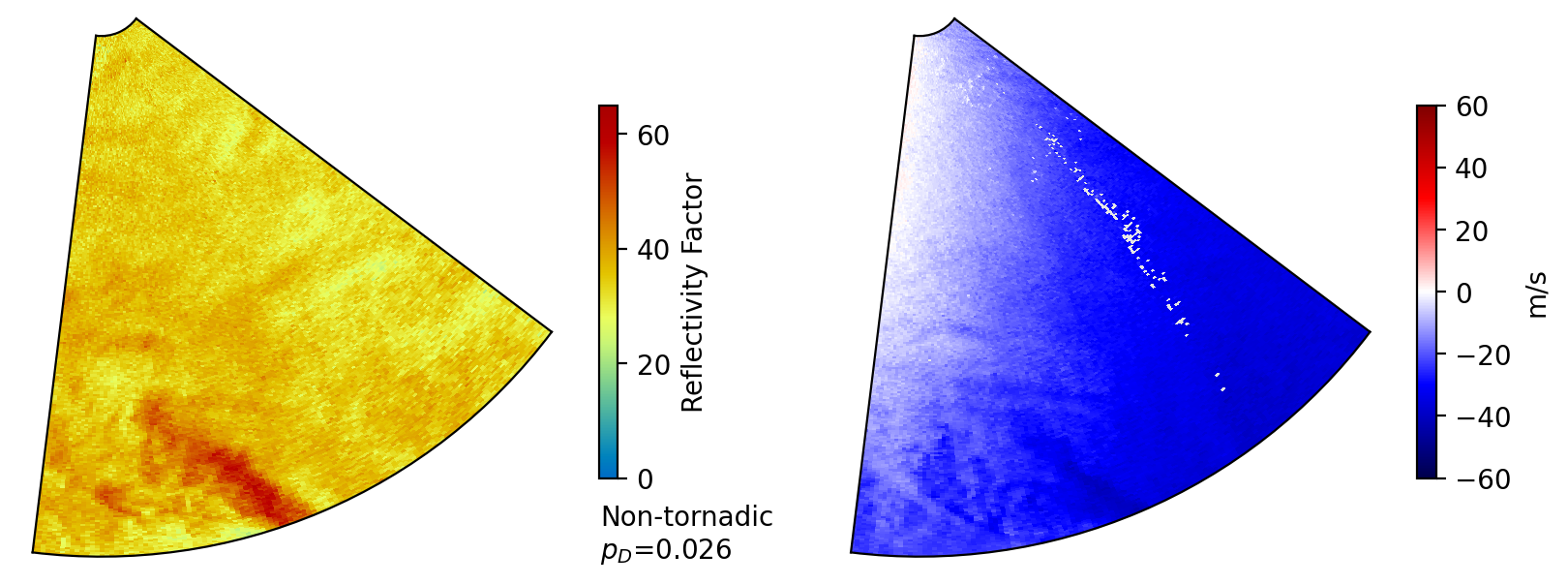}
\tcbincludegraphics{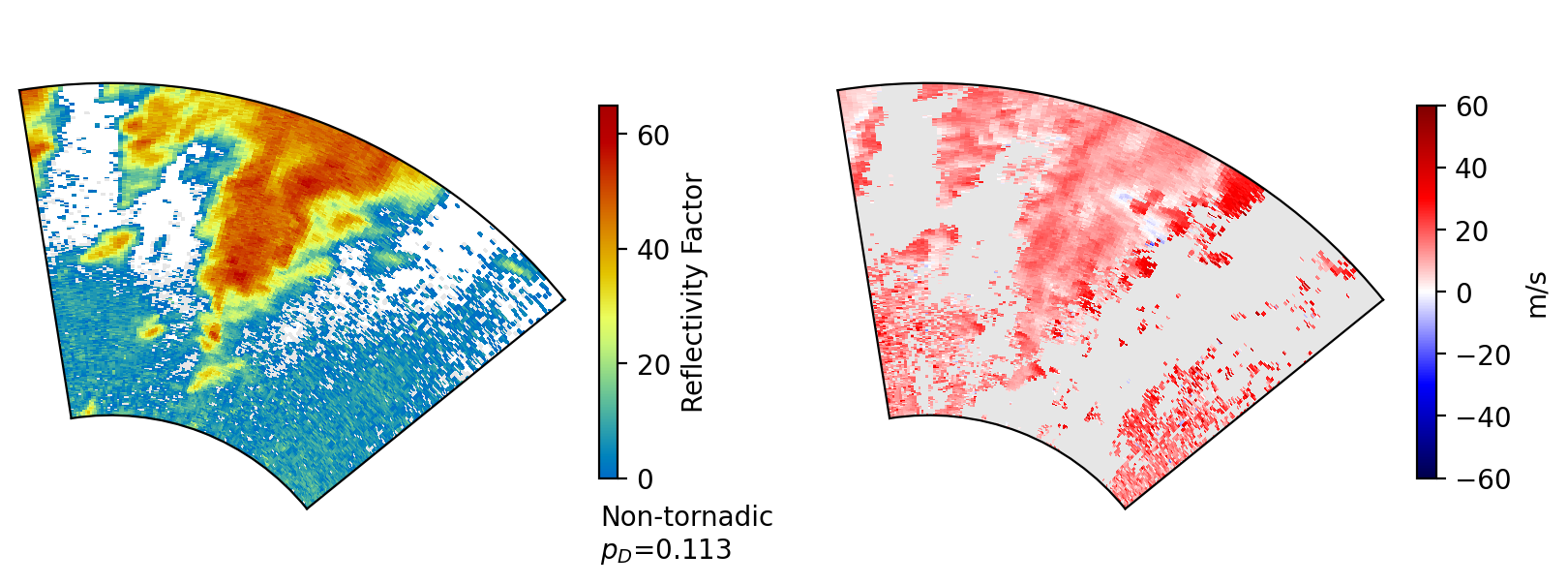}
\tcbincludegraphics{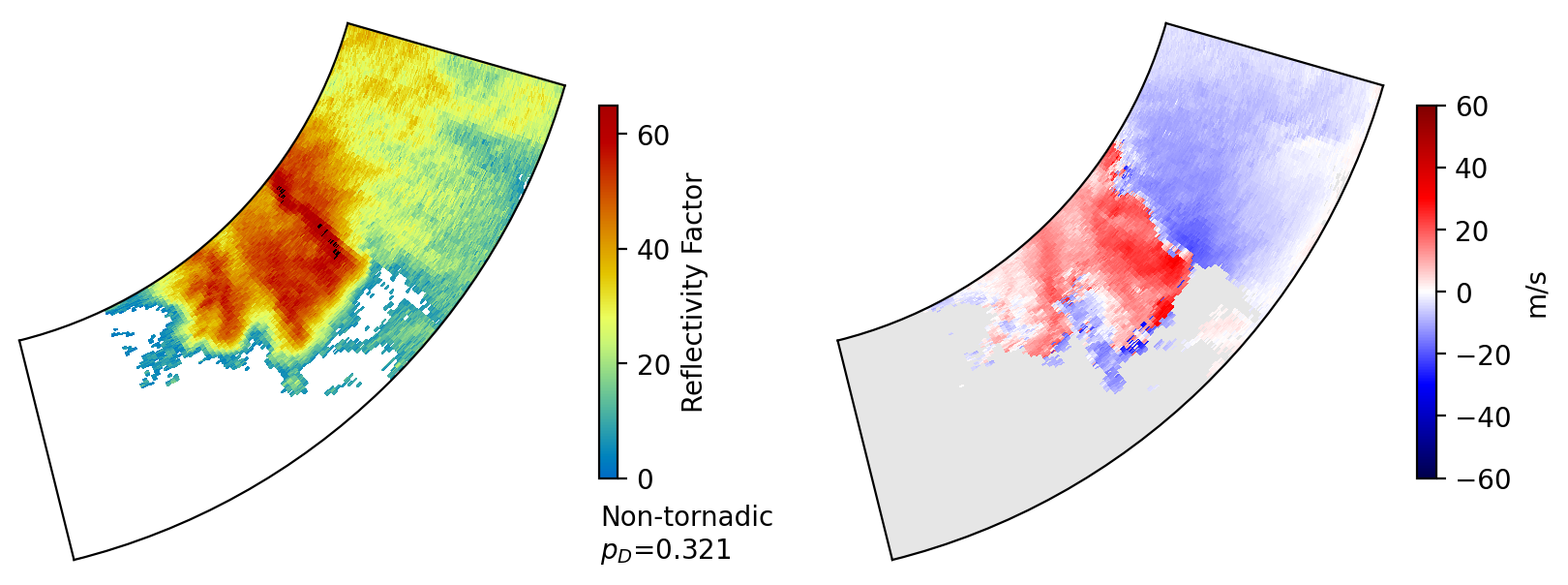}
\end{tcbraster}
\caption{\textbf{Correct Rejections}: Example non-tornadic cases that the CNN model correctly classified.}\label{f:cr}
\end{figure}

The hits shown in Fig. \ref{f:hits} show four samples that were correctly classified.  Each of these show common radar characteristics of tornado signatures, including hook echoes in the reflectivity factor field and velocity couplets in the radial velocity field.  Subjectively, the probabilities assigned with these samples seem to align well with the strength of the radar signals.   The correct rejections in Fig. \ref{f:cr} also demonstrate examples of obvious non-tornadoes with low assigned probability values.  It is interesting to note in the bottom right panel, the model gave a higher $p_D$ of 30\%, likely due to the strong line of shear observed in radial velocity, and the high reflectivity factor values near the edge of the storm.

\begin{figure}[h]
\centering
\begin{tcbraster}[raster columns=2,
    raster rows=2,
    colframe=black,
    raster equal height,
    boxsep=0pt,
    left=0pt,
    right=0pt,
    top=0pt,
    bottom=0pt,
    hbox,
    left skip=0pt,
    right skip=0pt,
  boxrule=0.5pt, 
    ]
\tcbincludegraphics{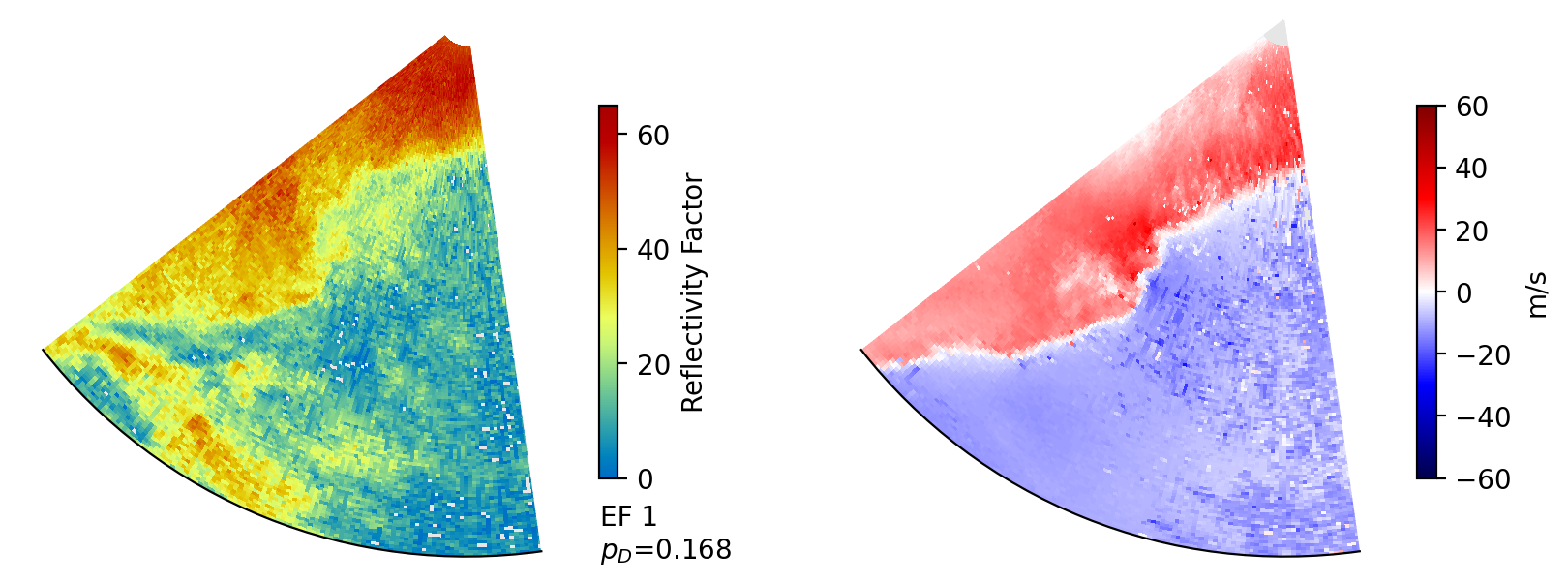}
\tcbincludegraphics{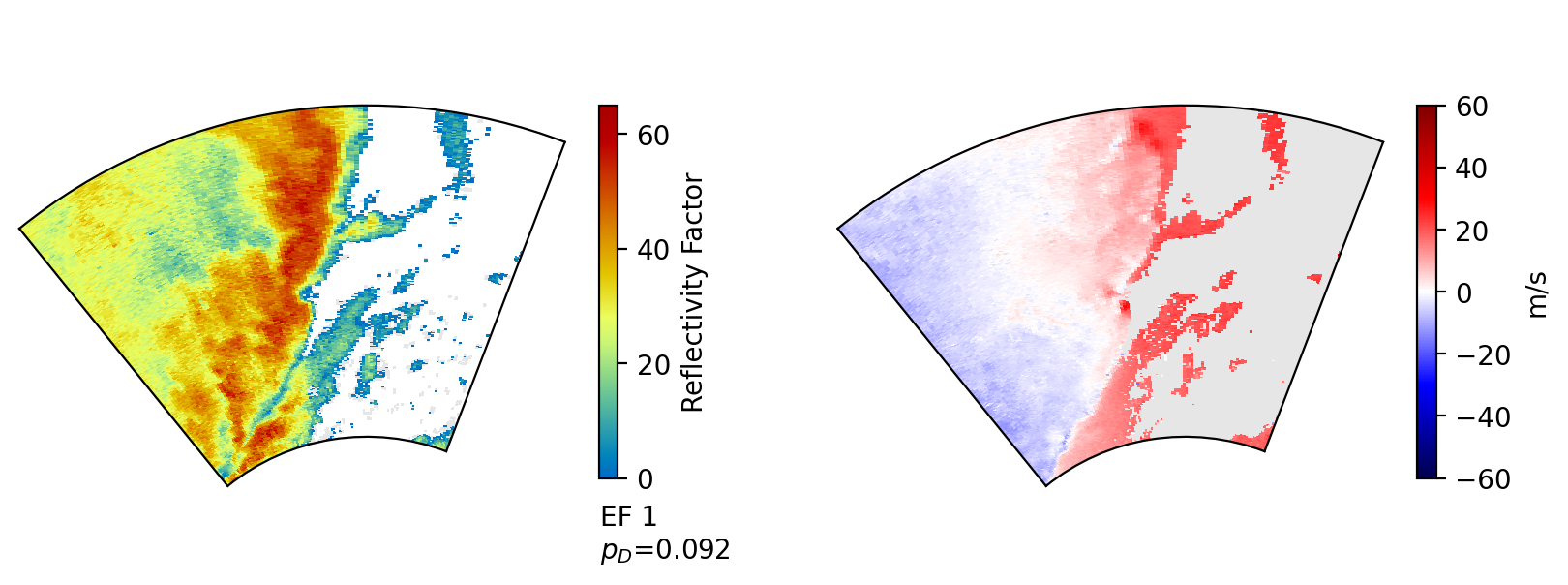}
\tcbincludegraphics{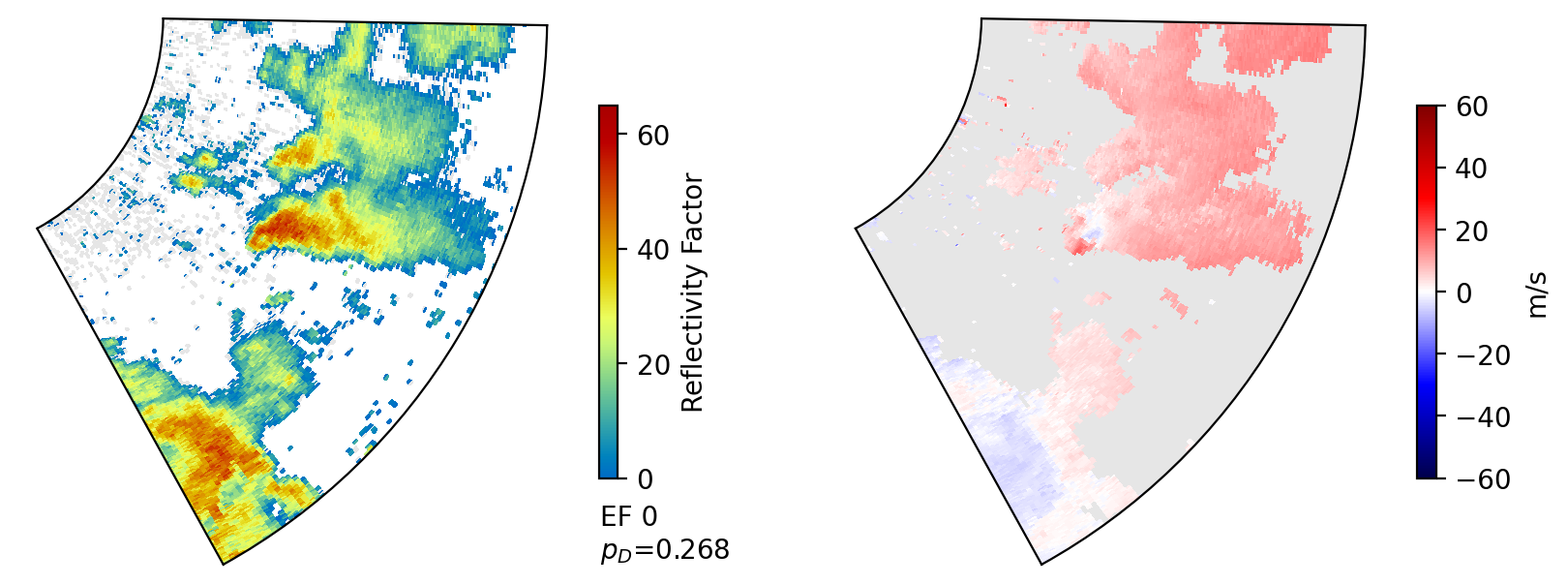}
\tcbincludegraphics{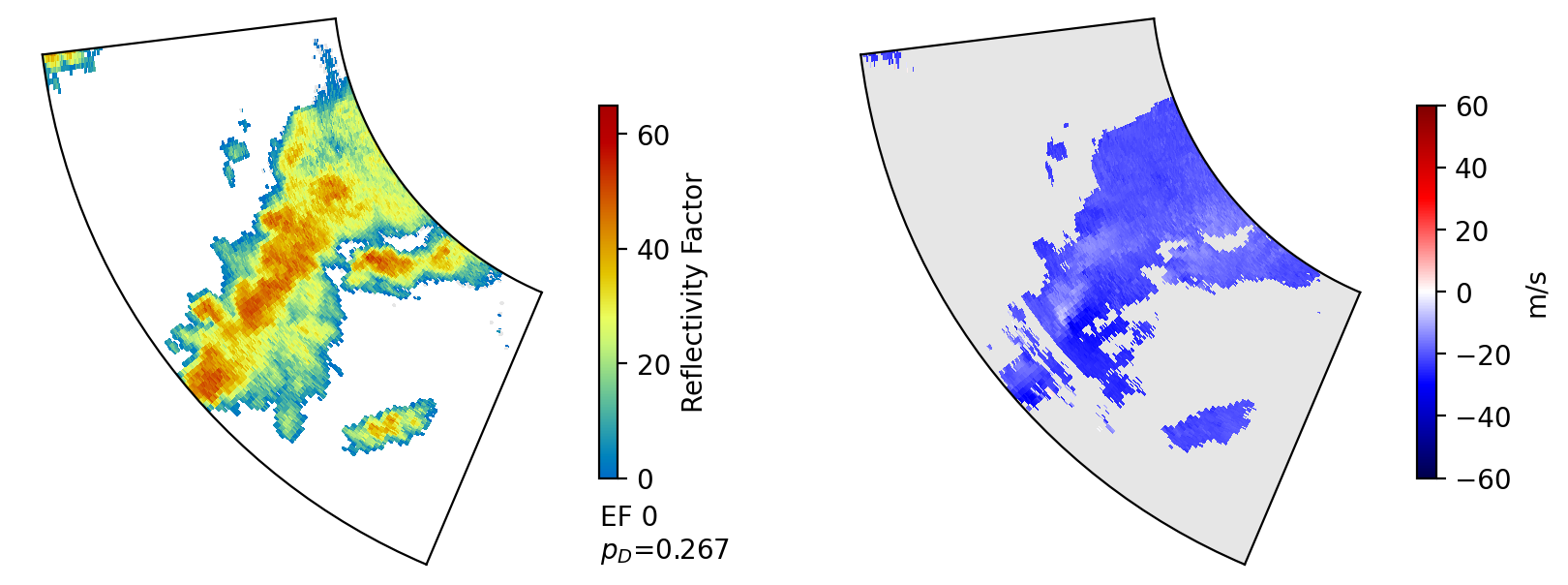}
\end{tcbraster}
\caption{\textbf{Misses}: Example tornadic cases that the CNN model incorrectly classified as non-tornadic.}\label{f:misses}
\end{figure}

\begin{figure}[h]
\centering
\begin{tcbraster}[raster columns=2,
    raster rows=2,
    colframe=black,
    raster equal height,
    boxsep=0pt,
    left=0pt,
    right=0pt,
    top=0pt,
    bottom=0pt,
    hbox,
    left skip=0pt,
    right skip=0pt,
  boxrule=0.5pt, 
    ]
\tcbincludegraphics{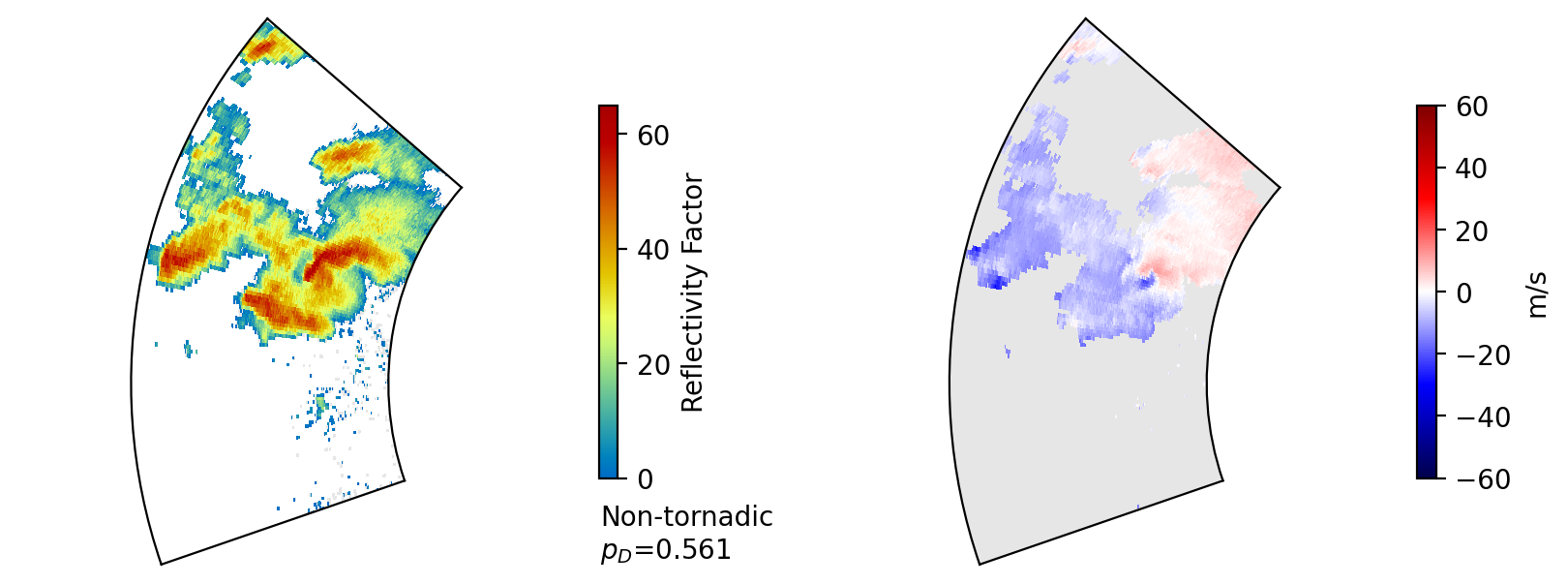}
\tcbincludegraphics{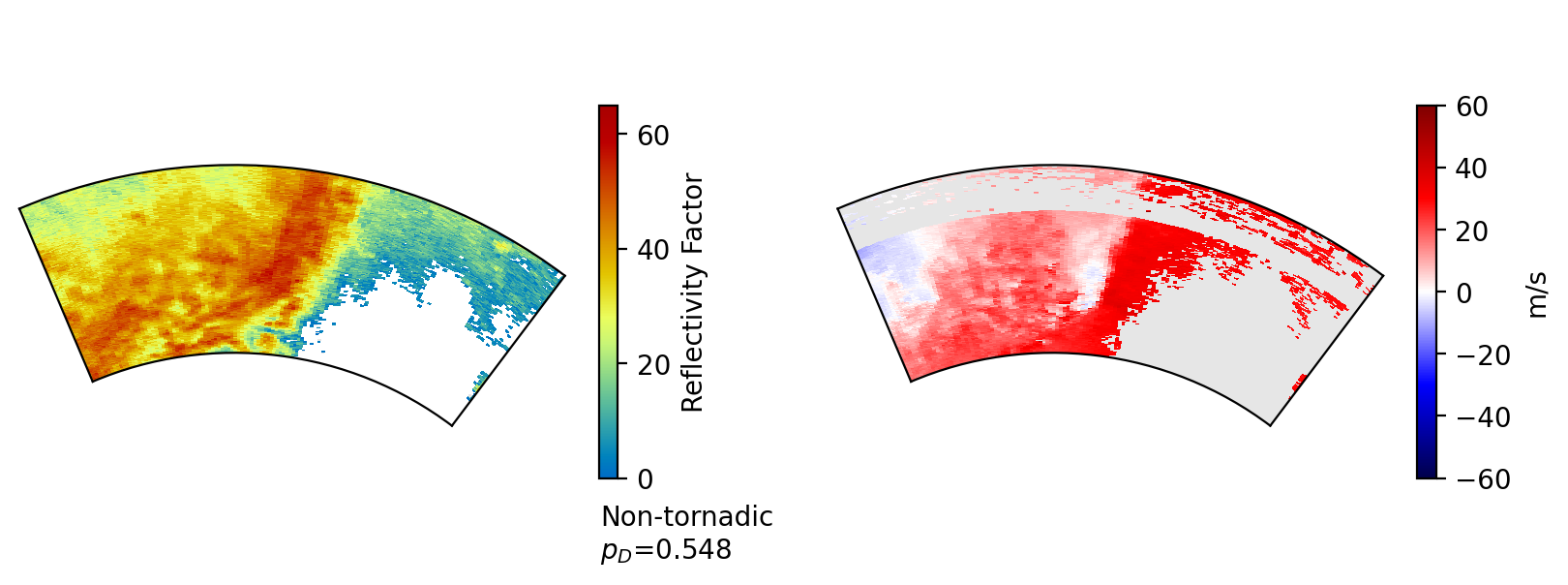}
\tcbincludegraphics{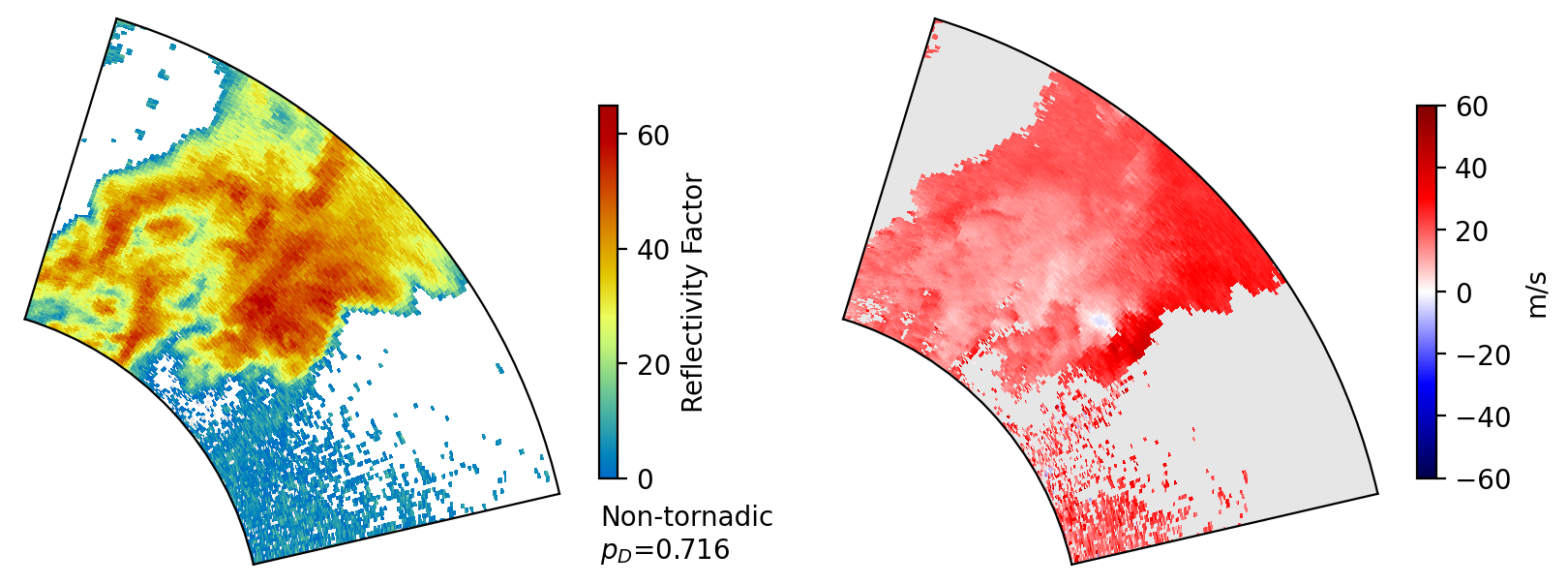}
\tcbincludegraphics{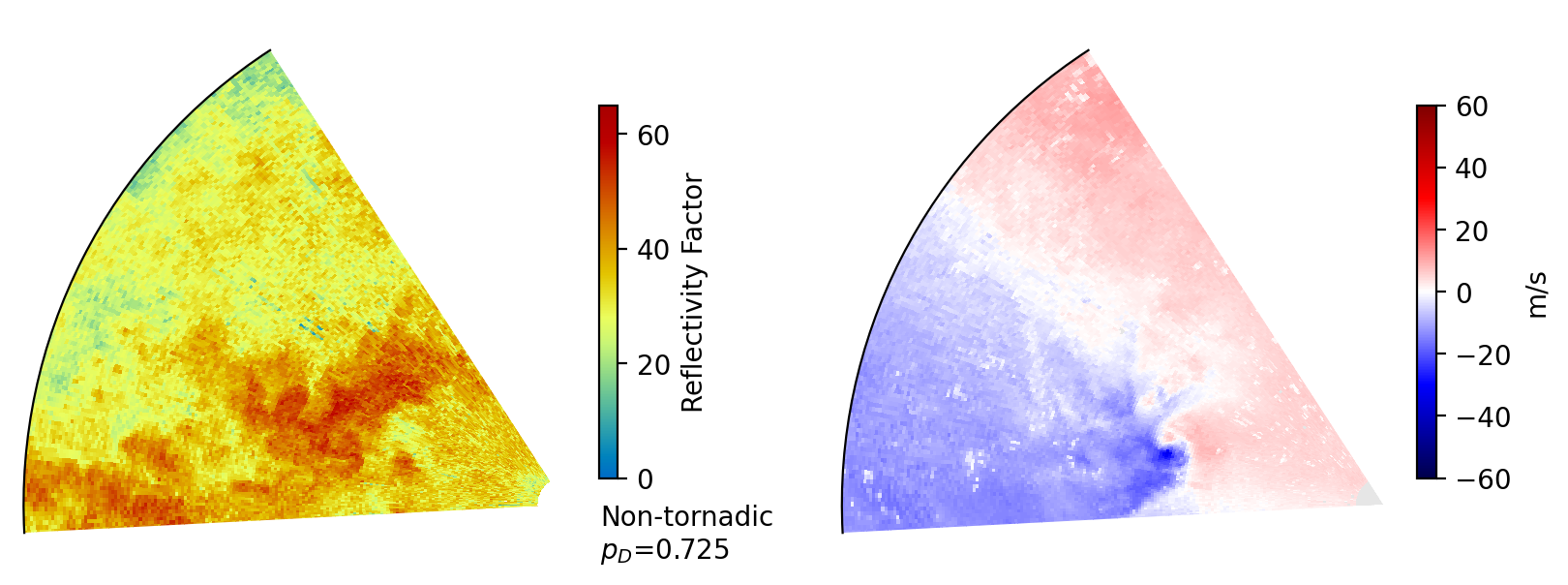}
\end{tcbraster}
\caption{\textbf{False Alarms}: Example non-tornadic cases that the CNN model incorrectly classified as tornadic.}\label{f:fas}
\end{figure}

The misses and false alarms in Figs. \ref{f:misses} and \ref{f:fas} depict incorrect classifications.  The misses shown are EF-0 and EF-1 tornadoes that do not exhibit extremely strong or obvious hook echoes or rotation signatures.  In the cases along the bottom of Fig. \ref{f:misses}, the CNN did assign low probabilities likely because of weak rotation signatures and/or presence of range folding (which would mask signal and increase uncertainty).   Each of the selected false alarm cases contains a clear region of rotation that was incorrectly classified as tornadic (albeit with moderate confidence of 50-75\%).  These appear to be rotating cells that lack other features that would increase the likelihood of being classified as tornadic.  

\section{Detection in Full Radar Scans}\label{s:case_studies}

This section demonstrates how ML models trained using \itorpe to classify tornado signatures in individual samples can be adapted for real-time monitoring of tornadoes on full-scale WSR-88D data.  While any of the ML baselines described in Section \ref{s:ml} could be adapted for this, the CNN model is highlighted in this section not only because it shows the highest overall skill, but also because CNNs are extremely efficient at processing large imagery, with inference on a full WSR-88D scan taking approximately one second on CPU, and fractions of a second on GPU.  In order to adapt the CNN for inference on full-sized imagery, the Global MaxPooling layer that computes the maximum likelihood over the entire chip is removed.  With this final layer removed, the CNN is fully convolutional (meaning it can be applied to input images of arbitrary size) and outputs a single-channel likelihood field that is sized proportionally to the input grid.  This output can be downscaled to the input image size using bilinear interpolation and resampled to Cartesian coordinates to produce a map of tornado likelihoods.  This section provides demonstrations of how the trained CNN model performs on entire storm episodes in the \itorpe test set.

The first case study considered is a tornado outbreak that occurred in the Lower Mississippi (MS) River and TN River Valleys on 28 April 2014 starting at approximately 1800 UTC, with Storm Episode ID 84120.  The storm system, driven by a strong jet stream and deep surface cyclone, caused damaging wind, large hail, and multiple tornadoes across the NWS Jackson, MS forecast area. The most destructive tornado was an EF-4, covering 34.3 miles and causing 10 fatalities. In total, 21 tornadoes were confirmed in the region, ranging from EF-0 to EF-4, with several counties experiencing significant damage.  The confirmed tornadoes from this outbreak resulted in 70 \itorpe radar image samples.

\begin{figure}[h]
\centering
  \noindent\includegraphics[width=35pc,angle=0]{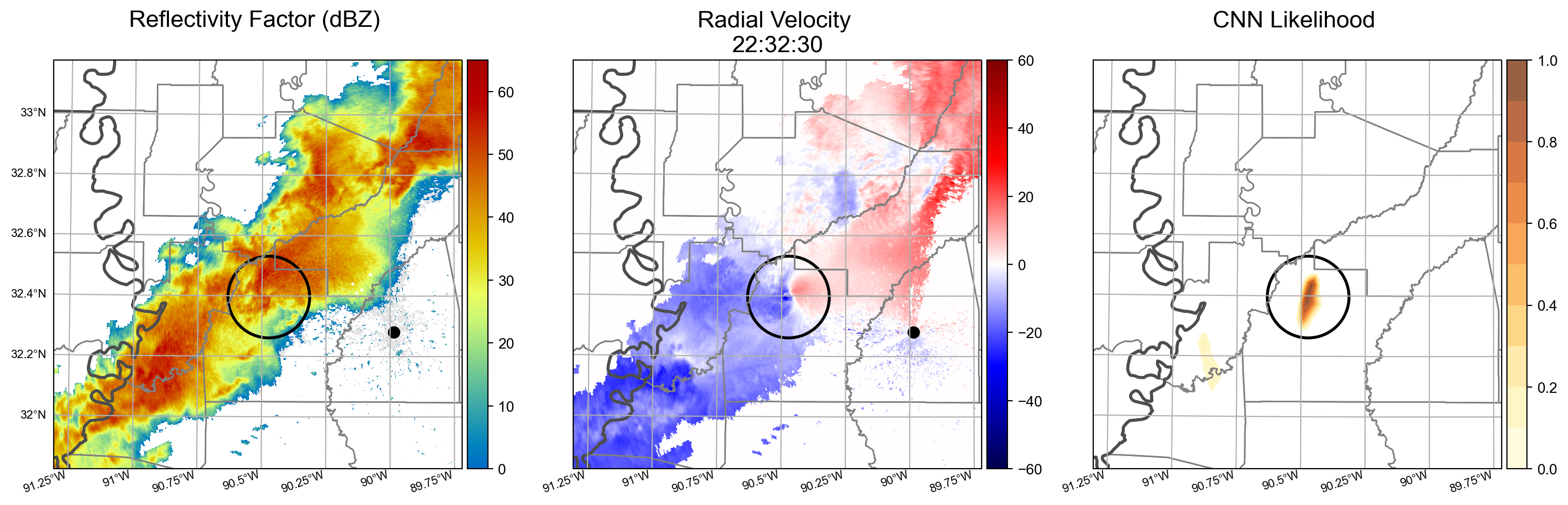}\\
  \caption{ Detection of a tornado near KDGX on 28 April 2014 using the CNN model trained on TorNet.  Reflectivity factor and radial velocity are shown in the left and middle panels, respectively.  The circle in the middle of the images indicates the location of a confirmed tornado. The CNN model described in Section \ref{s:ml} was applied to the entire radar scene, and the associated likelihood field is shown in the right panel, with a strongly elevated signal corresponding to the location of the tornado and low likelihood in areas away from the tornado.     }\label{fig:kdgx_sample}
\end{figure}

A single timestamp from this episode is shown in Fig. \ref{fig:kdgx_sample}.  The left two panels depict DBZ and VEL, respectively, at 2232:30 UTC on 28 April 2014.  At this time, a strong tornado signature emphasized by a hook echo in reflectivity factor and a strong couplet in radial velocity, indicated within the black circle.  The right-most panel provides output of the CNN model, which creates a likelihood at each grid cell.  In this case, the CNN produces a high likelihood that lines up with the features observed in reflectivity factor and radial velocity.  Moreover, the signal is localized in the region around the tornado, with no other high likelihoods present in other areas of the scene.  

In order to visualize performance over an entire storm episode, the likelihood shown in the right panel of Fig. \ref{fig:kdgx_sample} can be aggregated by computing the maximum observed likelihood over all WSR-88D observations in either time or space.  For example, when spatial likelihood maps are aggregated over time, they depict ``tracks'' of high likelihood that can be compared to confirmed tornado track(s) (black lines), similar to to Azimuthal Shear is aggregated in \citep{mahalik2019estimates}.  This is a useful spatial verification technique to see identified tornadoes, misses, and/or false alarms.  Alternatively, the likelihoods for each time can be aggregated over a region of interest to create time series.

\begin{figure}[h]
\centering
  \noindent\includegraphics[width=35pc,angle=0]{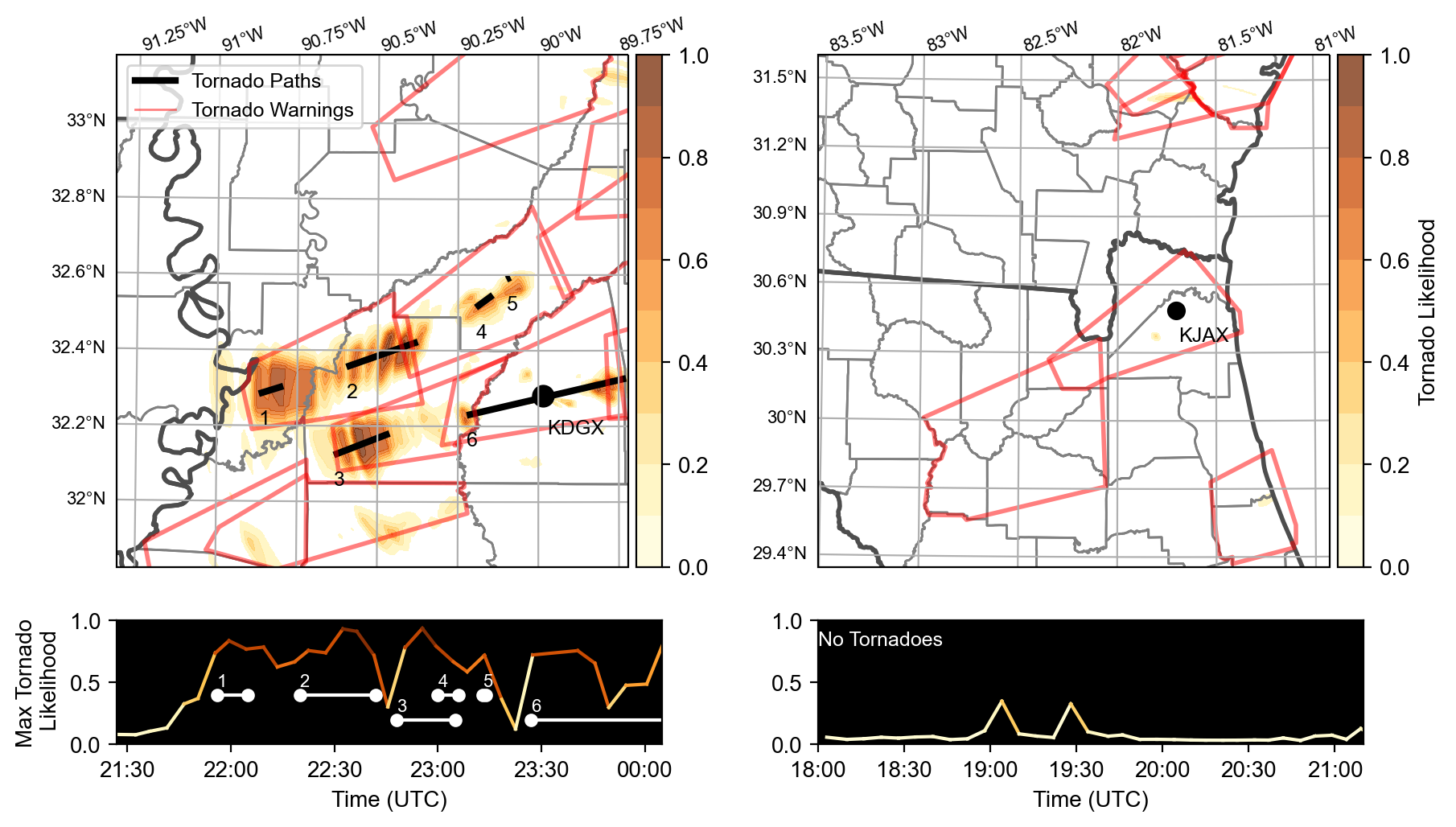}\\
  \caption{  Examples of tornado likelihoods aggregated over two selected storm episodes. (left) A tornado outbreak starting 28 April 2014 near KDGX.   The black lines represent confirmed tornado tracks.  CNN tornado likelihoods were aggregated over a three-hour period and are shown as shaded areas.  The maximum likelihood aggregated over the region for each time is plotted along the bottom.  In this case, the tracks of high likelihood correspond to the confirmed tornado tracks. (right) A similar visualization starting 3 March 2018 near KJAX with no confirmed tornadoes.  In this case, the likelihood remains low (<50\%) for the entirety of the time interval considered.  Tornado warnings were issued for this case, and they are shown as red polygons. }\label{fig:cumulative_likelihoods}
\end{figure}

For example, aggregated likelihoods computed between 2127 UTC 28 April to 0019 UTC 29 April are shown in the left panel of Fig. \ref{fig:cumulative_likelihoods}.   All confirmed tornadoes within this time range are denoted with black lines.  Most of the tornado tracks in this time range correspond with a region of high likelihood, with weaker signals being seen in the right-most track corresponding to a tornadic cell that passed directly over the radar.  This situation likely created difficulty for the CNN algorithm, which is limited by kernel sizes of the CNN kernels and thus was not able to ``see'' the entire tornado represented in polar coordinates (a limitation of this approach when very close to the radar).  Moreover, regions not near a tornado track exhibit lower likelihoods (<50\%).  The red polygons correspond to tornado warnings issued during this time period.  Note that regions of low-moderate likelihoods in the southern part of the image are contained in these polygons.  

The right panel of Figure \ref{fig:cumulative_likelihoods} shows another storm episode near KJAX starting 3 March 2018.  Unlike the previous case, this episode did not exhibit any confirmed tornadoes, although tornado warnings were issued.  In this case, the aggregated likelihood map shows overall low probabilities (<50\%) everywhere in the region, with the only occurrences happening in areas where tornado warnings were issued.

\section{Conclusion}

This study introduces a public benchmark dataset called \itorpe for tornado detection and prediction.  Benchmark datasets are a critical component of AI/ML fields, and meteorology is no different.  A combination of data sources, including the NCEI storm events database and Level-II/III WSR-88D data at full resolution and multiple tilts, are used to create hundreds of thousands of samples focused on relevant tornadic, and non-tornadic scenes.  This dataset was curated and labeled for ML purposes, although such a dataset may be useful in many other ways, including large-scale case studies, automation, and algorithm development and testing.  The dataset is made freely available in a format that can be easily parsed and filtered, and is straightforward to work with in a users programming environment of choice.

A number of ML baselines were developed and compared for the important problem of tornado detection.
Baseline results showed that performance of ML models trained on \itorpe exceed that of an operational baseline (TVS).  The best performing (and most complex) model was a CNN, a DL model that utilized full resolution imagery directly and did not need features to be defined and extracted beforehand.  This model was also demonstrated on selected case studies to show how ML models trained on \itorpe can be adopted for real-time detection and tracking of tornadoes in full-sized radar imagery. 

It is hoped that in future studies, the dataset leads to innovation in this research area that will  exceed and/or build upon the baselines presented in this work.  Future directions include the possible addition of more data sources,  such as more tilts/time frames of radar data, various satellite imaging channels, lightning from the global lightning mapper (GLM) aboard the Geostationary Operational Environmental Satellite (GOES) constellation, and NWP model output.  Multiple studies show benefits of ML for fusing multiple modalities of data, leading to our hope that continuing to build \itorpe will yield increasingly interesting and impressive results.  Taking the need to develop this dataset from scratch out of the time budget for researchers is anticipated to allow for new and improved AI/ML techniques for this problem, including those who may not be domain experts.  Additionally, through the public release and open-source nature of the dataset, it is hoped that continued growth and improvement driven by the community will only strengthen its impact.  We hope to see this approach taken in many contexts within the field of meteorology in the coming years.

Going forward with this specific dataset, the problem of tornado \textit{prediction} is an obvious next step.  As mentioned previously, the \itorpe dataset contains data before tornadogenesis, allowing for the potential to predict tornadoes at given temporal thresholds.  While the utility of this dataset to the prediction problem has not yet been fully explored, it is the immediate next step for the ongoing research.  When combined with additional data modalities, it is hoped that viability for a tornado prediction algorithm will be sufficient.

%

\clearpage
\acknowledgments
DISTRIBUTION STATEMENT A. Approved for public release: distribution unlimited. This material is based upon work supported by the Department of the Air Force under Air Force Contract No. FA8702-15-D-0001. Any opinions, findings, conclusions or recommendations expressed in this material are those of the author(s) and do not necessarily reflect the views of the Department of the Air Force.

%
%
\datastatement
Instructions for downloading the \itorpe dataset, as well as tools for utilizing \itorpe in ML pipelines can be found at https://github.com/mit-ll/tornet.  Archives of Level-II WSR-88D data used for building \itorpe can be found from AWS Open Data Registry https://registry.opendata.aws/noaa-nexrad/.  Level-III archives for the WSR-88D network were obtained from Google Cloud Storage: https://console.cloud.google.com/storage/browser/gcp-public-data-nexrad-l3.   The NSED can be downloaded from NCDC at https://www.ncdc.noaa.gov/stormevents/.  Tornado warning polygons can be obtained from Iowa State University's archive found at https://mesonet.agron.iastate.edu/vtec/search.php.








%



\appendix[A] 

\appendixtitle{Feature Extraction}

\begin{figure}[h]
  \noindent\includegraphics[width=35pc,angle=0]{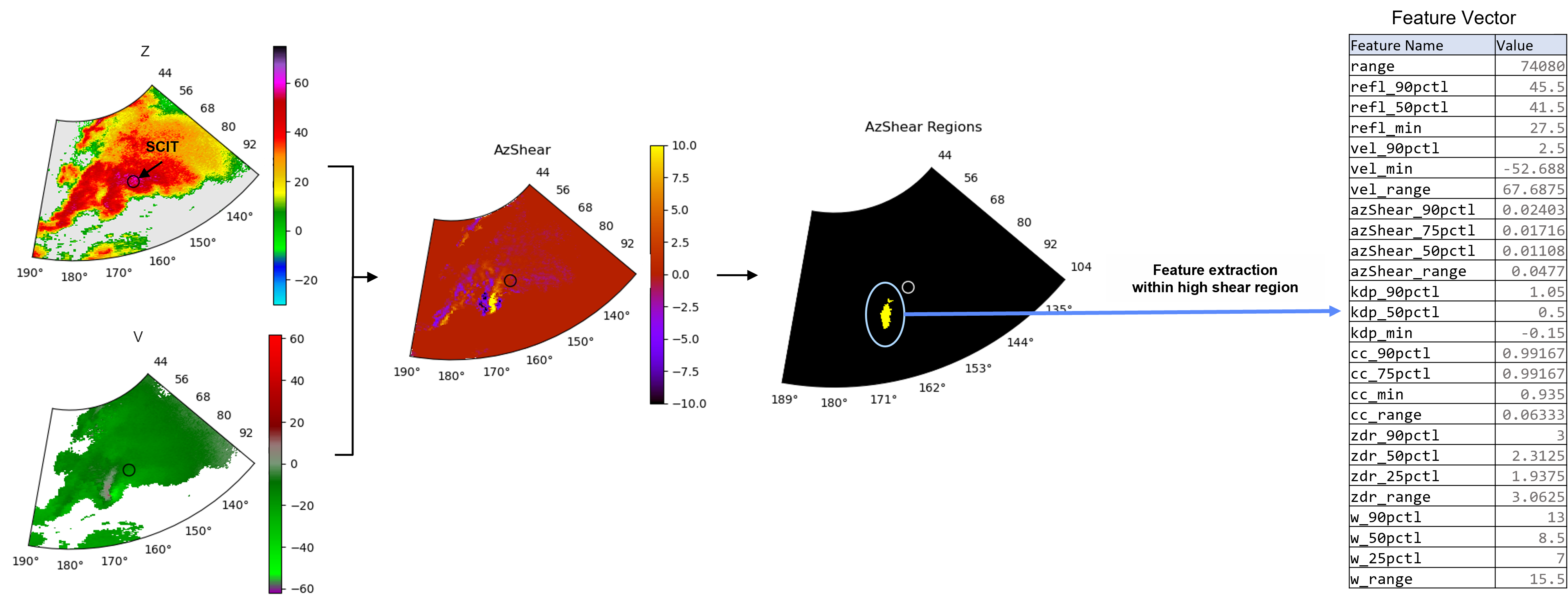}\\
  \caption{ Process for extracting features from data chips for use in non-DL models.  Tilts of REF and VEL (left images) are combined to compute azimuthal shear (second image from the left).  Regions of high shear are identified using an object identification algorithm (third image from left).   Within these identified objects, a feature vector is computed across a number of different radar values.}\label{fig:featureextract}
\end{figure}

This section describes the feature extraction process for non-DL methods covered in Section \ref{s:ml}\ref{ss:baselines}.  An overview of feature extraction is shown in Fig. \ref{fig:featureextract}.   First, the reflectivity factor and radial velocity channels are combined to create an azimuthal shear field (AzShear).   AzShear is created by first isolating regions with reflectivity above a threshold of 20 dBZ.  The azimuthal gradient of radial velocity is computed using the Linear Least Squares Derivative method \citep[LLSD;][]{smith2004use}.  Next, a threshold of 0.006 $s^{-1}$ is applied to the AzShear field, and a region extraction algorithm is applied to isolate regions of high AzShear larger than a minimal size threshold.  If any regions remain, the one within the chip containing the highest level of AzShear is selected.  If no regions remain, then a circular region in the center of the chip is used instead.  

To extract a feature vector, a number of statistics listed in the table in the right portion of Fig. \ref{fig:featureextract} are measured from within the selected region of high AzShear across the six radar variables from each \itorpe sample, and the derived AzShare field.  Statistics included for each variable include the 90th percentile, 75th percentile, 50th percentile, 25th percentile, minimum value, and range of values.  In addition, the chip's range from the radar is also included as a feature.  Only the lowest tilt was used for feature extraction.

\appendixtitle{Additional Performance Curves}\label{s:other_roc}

In addition to the curves shown in Fig. \ref{fig:curves_all}, ROC curves and PD were also computed for two additional views of the test sets (ii) and (iii) described in Section \ref{ss:test_performance}.  

\begin{figure}[h]
  \noindent\includegraphics[width=35pc,angle=0]{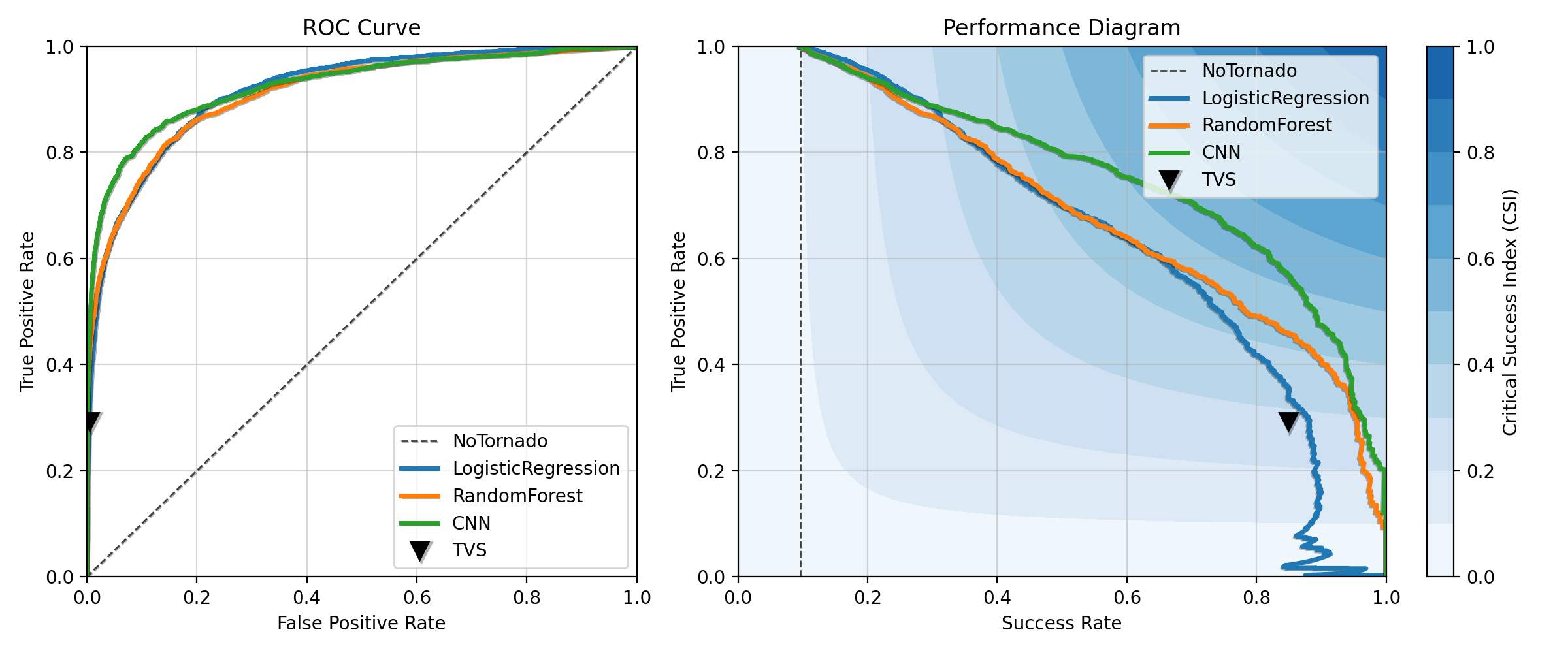}\\
  \caption{ Same as Fig. \ref{fig:curves_all}, except only the ``random'' category is used for non-tornadoes. }\label{fig:curves_random}
\end{figure}

Results for test partition (ii), where only the random category are used for null cases, is shown in Fig. \ref{fig:curves_random}.  Compared to the results in Fig. \ref{fig:curves_all}, it is clear that performance increases across all models.   This is due to the fact that warning samples contain cases that are harder for the models to classify correctly.  With this partition, the logistic regression and random Forest models show similar performance, and still under-perform the CNN for most thresholds.  The TVS again shows a low FPR / low TPR, but still shows worse performance than the ML counterparts.

\begin{figure}[h]
  \noindent\includegraphics[width=35pc,angle=0]{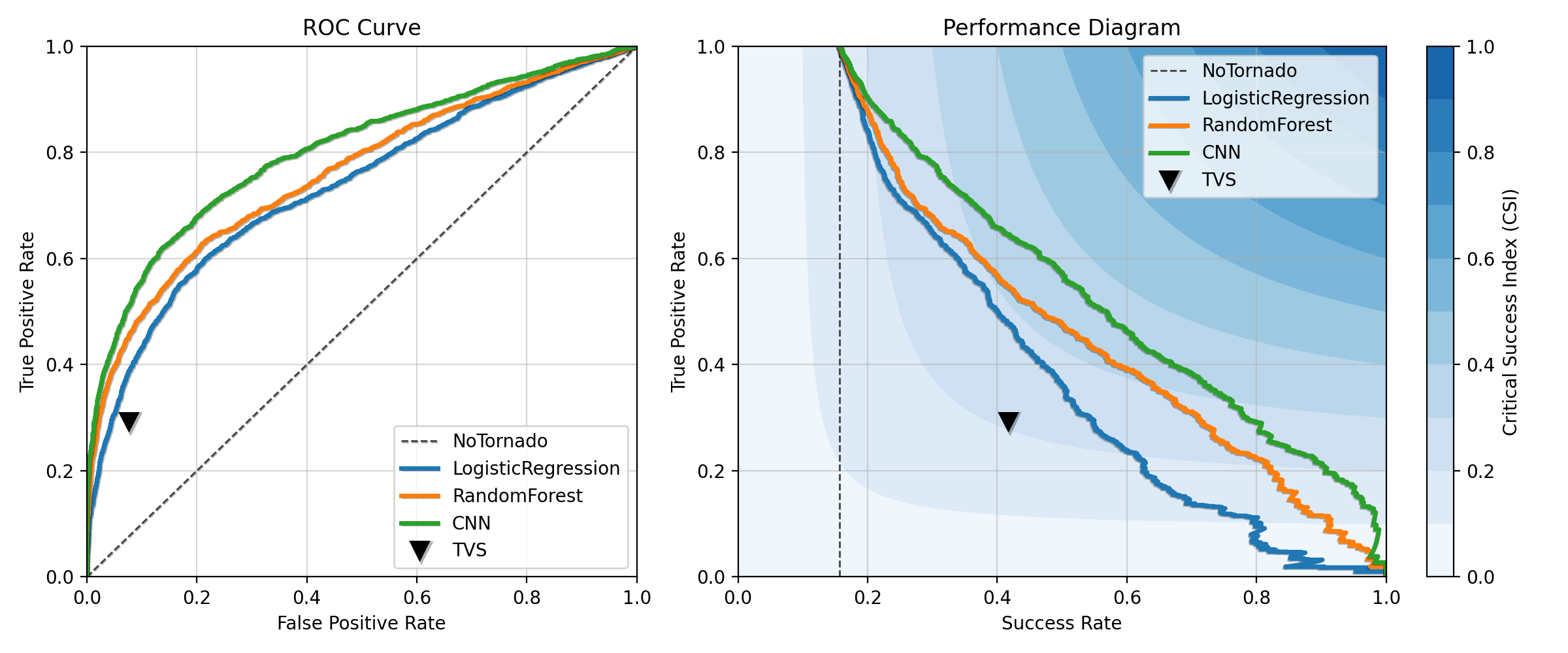}\\
  \caption{ Same as Fig. \ref{fig:curves_all}, except only the ``warnings'' category is used for non-tornadoes. }\label{fig:curves_warnings}
\end{figure}

Results for the final partition (iii) are shown in Fig. \ref{fig:curves_warnings}.   Compared to the ROC curves in partitions (i) and (ii), results for this partition are the worst overall.  This is to be expected, since the warning categories contain imagery that led human forecasters to issue tornado warnings.  Not surprisingly, this results in the classification models to all have higher FPR compared to the other partitions.  Also of note, is that the PD bears similarity to partition (i), implying that most of the false positives (which drives SR) observed in partition (i) came from the warnings category.  Within this partition, we again see the CNN outperforming other baselines, and TVS under-performing all ML baselines considered.

\bibliographystyle{ametsocV6}
\bibliography{Kurdzo_Bibliography,references.bib}

\end{document}